\documentclass[twocolumn,aps,prl,superscriptaddress]{revtex4-2}

\usepackage{graphicx}% Include figure files
\usepackage{gensymb}
\usepackage{dcolumn}% Align table columns on decimal point
\usepackage{bm}% bold math
\usepackage{textcomp}
\usepackage[intlimits]{amsmath}
\usepackage{esint}
\usepackage{braket}
\usepackage{color}
\usepackage{bbold}
\usepackage{soul} %for \hl and \st
\usepackage{titlesec}

\titleformat{\section}
  {\normalfont\large\bfseries\raggedright
   \hyphenpenalty=10000\exhyphenpenalty=10000}
  {}
  {0pt}
  {}

\titleformat{\subsection}
  {\normalfont\normalsize\bfseries\raggedright
   \hyphenpenalty=10000\exhyphenpenalty=10000}
  {}
  {0pt}
  {}

\begin{document}
	
\title{Largest eigenvalue statistics of wavefront shaping in complex scattering media}

\author{Gr\'egory Schehr}
%\email{schehr@lpthe.jussieu.fr}
\affiliation{Sorbonne Universit\'e, Laboratoire de Physique Th\'eorique et Hautes Energies CNRS UMR 7589, 4 Place Jussieu, 75252 Paris Cedex 05, France}%
    
\author{Hasan Y{\i}lmaz}
%\email{hasan.yilmaz@slu.edu}
\affiliation{Department of Electrical and Computer Engineering, Saint Louis University, St. Louis, Missouri 63103, USA}

\begin{abstract}
\noindent
Corresponding authors: G.S. (schehr@lpthe.jussieu.fr) and
H.Y. (hasan.yilmaz@slu.edu).

\vspace{0.5em}

In wavefront shaping, light, sound, and other waves are focused through complex scattering media onto one or more target positions, and the resulting intensity enhancement is quantified by the enhancement factor. While reproducible enhancement is crucial in experiments, the fluctuations of the enhancement factor remain largely unexplored. Here, we combine experiments, numerical simulations, and exact random-matrix theory to determine its full distribution for multi-target focusing. Exact finite-size random-matrix predictions accurately describe both the mean enhancement factor and its fluctuations beyond the asymptotic Marčenko–Pastur regime, whenever long-range mesoscopic correlations are negligible (e.g., in weakly scattering media or when only a limited number of input channels is controlled). In contrast, strongly scattering media exhibit giant enhancement-factor fluctuations that increasingly exceed these parameter-free predictions as the number of controlled input channels increases. These findings establish the enhancement factor not only as a measure of focusing performance, but also as a sensitive statistical observable that provides a simple and experimentally accessible probe of long-range mesoscopic correlations.

\end{abstract}
	
\maketitle
	
\section{Introduction}

Wavefront shaping stands as a paradigm-shifting framework for the coherent control of light, sound, and other waves in complex scattering media, such as biological tissue and disordered materials~\cite{2007_Vellekoop, 2012_Mosk_NatPhoton_R, 2015_Park_R, 2015_Vellekoop_OptExpress, 2015_Yang_NatPhoton, 2017_Rotter_RMP_R, gigan2022roadmap, cao2022shaping}. Experimental demonstrations have shown that by precisely tailoring the incident wavefront, the total transmission of microwaves, acoustic, elastic, or optical waves can be dramatically enhanced—or waves can even be tightly focused—through strongly scattering media~\cite{lerosey2007focusing, 2007_Vellekoop, 2008_Vellekoop_PRL, 2010_vellekoop_Nat._Photonics, davy2012focusing, 2012_Choi_Nat._Photonics, davy2013transmission, 2014_Popoff_PRL, ojambati2016coupling, 2017_Wade_Nat._Phys., 2019_Yilmaz_Nat._Photonics, 2019_Yilmaz2019_PRL}. In optics, this level of precise coherent control over light holds profound implications for imaging~\cite{2010_vellekoop_Nat._Photonics, 2010_Vellekoop_OL, 2010_Psaltis_OE, 2011_VanPutten_PRL, 2012_Katz_NatPhoton}, quantum secure optical authentication~\cite{goorden2014quantum}, high-dimensional quantum optics~\cite{Huisman2014, Huisman2015, wolterink2016programmable}, optical manipulation~\cite{Peng2019}, photothermal therapy~\cite{2015_Park_R, wang2015controlled}, microsurgery~\cite{yanik2004functional}, and optogenetics in complex scattering environments~\cite{fenno2011development, ruan2017deep, pegard2017three}, including biological tissue, ensembles of micro- and nanoparticles, and other seemingly opaque media.

\begin{figure}[th]
\centering
\includegraphics[width=\linewidth]{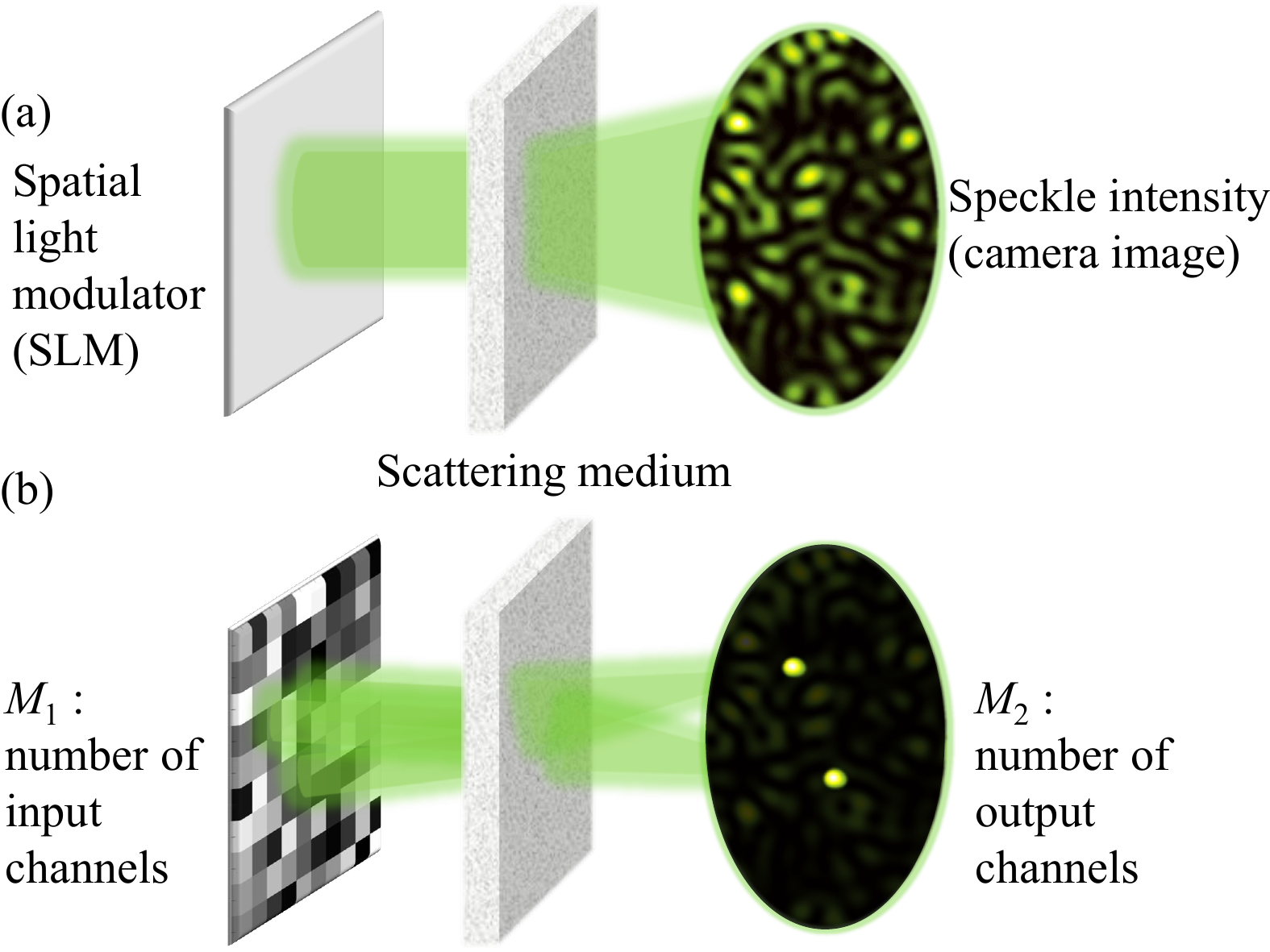}
\caption{The wavefront-shaping concept to focus light through a scattering medium onto multiple target points is shown. (a) A sketch of the speckle pattern in transmission when an unshaped wavefront is incident onto a scattering medium is demonstrated. (b) The incident wavefront is shaped by a spatial light modulator with $ M_1$ degrees of freedom at the input, focusing light onto two selected target points (output channels) $M_2 = 2$.}
\label{figure1}
\end{figure}

The measurement of the transmission matrix—encoding the input–output field relationship of a medium—is essential for controlling the output field via input wavefront shaping~\cite{2010_Popoff_PRL}. Once known, the transmission matrix enables focusing light to single or multiple target positions~\cite{prada1994eigenmodes, 2007_Vellekoop, 2008_Vellekoop_OE, 2010_Popoff_PRL, 2010_vellekoop_Nat._Photonics, 2010_Vellekoop_OL, 2010_Psaltis_OE, fenno2011development, 2011_VanPutten_PRL, 2012_Katz_NatPhoton, van2012nonimaging, goorden2014quantum, Huisman2014, Huisman2015, wolterink2016programmable, xu2017focusing, 2017_Wade_Nat._Phys., ruan2017deep, pegard2017three, Feng2019, Peng2019, Hu2023, hong2024robust, cheng2025binary}, or transmitting images through scattering media~\cite{popoff2010image}. In particular, simultaneous focusing on multiple positions has important applications in optical metrology~\cite{van2012nonimaging}, optical trapping of microparticles~\cite{Peng2019}, programmable multiport optical circuits~\cite{Huisman2014, Huisman2015, wolterink2016programmable}, and focusing on extended guide stars~\cite{2008_Vellekoop_OE, Zhou2014, ruan2017deep, pegard2017three}. A key metric for evaluating the efficacy of wavefront shaping in these contexts is the enhancement factor, which quantifies the intensity enhancement achieved at the target positions~\cite{2007_Vellekoop, 2008_Vellekoop_OE, 2015_Vellekoop_OptExpress}. It is defined as $\eta = \sum_{m = 1}^{M_2} |\sum_{n = 1}^{M_1} A_{mn}E_n^{\rm opt}|^2/\overline{\sum_{m = 1}^{M_2} |\sum_{n = 1}^{M_1} A_{mn}E_n}|^2$, where \(A\) is the \(M_2 \times M_1\) sub-transmission matrix of the medium, with $n$ labelling the incident Fourier components at the input and $m$ the position of the targets at the output. Here $M_1$ denotes the number of degrees of freedom at the input while \(M_2\) denotes the number of target (focusing) points, see Fig.~\ref{figure1}. The vector \(E^{\rm opt}\) is the input wavefront that maximizes the total intensity over these \(M_2\) targets. The denominator in the definition of $\eta$ is the same total intensity, but averaged over statistically equivalent disorder realizations (i.e., over different sub-transmission matrices \(A\)) or, equivalently, over independent unit-power input wavefronts \(E\) whose complex field components are independent and identically distributed (i.i.d.) circularly symmetric complex Gaussian entries.
    
One typically employs an optimization algorithm to determine the optimal input wavefront $E^{\rm opt}$ that maximizes the intensity at the target locations~\cite{vellekoop2008phase,2015_Vellekoop_OptExpress}. For a single target ($M_2=1$), digital optical phase conjugation attains the global maximum of the enhancement factor $\eta$~\cite{2015_Vellekoop_OptExpress}. For multiple targets ($M_2>1$), the global maximum of $\eta$ equals the largest eigenvalue $\lambda_{\rm max}$ of the focusing operator $A^\dagger A$, where $A$ denotes an $M_2 \times M_1$ sub-transmission matrix of the disordered medium, with $t$ the normalized transmission matrix obeying ${\rm Var}(t_{mn})=1/M_2$ (see e.g. \cite{2017_Wade_Nat._Phys.}).

In practical wavefront-shaping focusing experiments, one typically operates in the regime $M_1 \gg M_2$. The dependence of the mean enhancement $\langle \lambda_{\rm max} \rangle$ on the number of controlled degrees of freedom $M_1$ and output targets $M_2$ has been observed experimentally and is widely recognized in the wavefront shaping community to exhibit the scaling $\langle \lambda_{\rm max} \rangle \sim M_1 / M_2$~\cite{2008_Vellekoop_OE, Zhou2014, 2015_Vellekoop_OptExpress, ruan2017deep}. On the other hand, if one assumes that the entries of $A$ are independent complex Gaussian numbers, then the focusing operator is an (anti)-Laguerre-Wishart (LW) random matrix~\cite{beenakker1997random}. In this case, for large $M_1$ and $M_2$ with fixed ratio $M_1/M_2$, the mean largest eigenvalue $\langle \lambda_{\max} \rangle$ coincides with the upper edge of the Mar\v{c}enko--Pastur (MP) spectrum \cite{marvcenko1967distribution, mehta2004random,forrester2010log}:
\begin{align}
\langle\lambda_{\rm max}\rangle \, = \left(\sqrt{\frac{M_1}{M_2}}+1\right)^2.
\label{eq:1}
\end{align}
While the MP law predicts $\langle\lambda_{\max}\rangle \sim M_1/M_2$ in the limit $M_1 \gg M_2$, the applicability of this formula (\ref{eq:1}) in experimental situations where $M_1$ and $M_2$ are finite remains questionable.

Furthermore, predicting the sample-to-sample and target-to-target fluctuations of the enhancement factor, i.e., $\Delta \lambda_{\max} \equiv \sqrt{\langle \lambda_{\max}^2 \rangle - \langle \lambda_{\max}\rangle^2}$, is key to establishing the fundamental limits of experimental reproducibility. Yet, its full statistical distribution and fluctuations remain largely unexplored, apart from the simpler case of single-point focusing $(M_2 = 1)$, which can be analysed rather straightforwardly~\cite{PaniaguaDiaz2023}.
   
In this work, we employ a random matrix framework to describe the full distribution of the enhancement factor for focusing light onto multiple target locations ($M_2>1$) through diffusive media. Under standard assumptions, the focusing operators can be described by random matrices belonging to the so-called Laguerre--Wishart ensemble~\cite{beenakker1997random}. This framework accurately predicts both the mean enhancement factor and its fluctuations at finite $M_1$ and $M_2$, extending beyond the asymptotic Mar\v{c}enko--Pastur limit and providing a parameter-free baseline for identifying excess fluctuations arising from long-range mesoscopic correlations. It is known that while long-range mesoscopic correlations do not affect the mean enhancement factor for small values of $M_2$, they increase the background intensity, with this effect becoming experimentally observable for large values of $M_1$~\cite{2008_Vellekoop_PRL, ojambati2016controlling, 2017_Wade_Nat._Phys., shaughnessy2024multiregion, van2025mesoscopic}. These long-range mesoscopic correlations originate from coherent interference among multiple scattering paths and constitute one of the defining signatures of mesoscopic wave transport in disordered media~\cite{2007_Akkermans, beenakker1997random, berkovits1994correlations, van1999multiple, 2017_Rotter_RMP_R, 1987_Cwilich, 1988_Mello_Akkermans_Shapiro, 1988_Stone, 1989_Shapiro, 1990_Genack, 2013_Muskens}. They manifest themselves through connected four-field correlation functions and underlie several hallmark mesoscopic phenomena, including universal conductance fluctuations~\cite{imry1986active, scheffold1998universal}, the emergence of open transmission channels with near-unity transmittance~\cite{beenakker1997random, 2017_Rotter_RMP_R, cao2022shaping, sarma2016control, 2019_Yilmaz_Nat._Photonics, 2019_Yilmaz2019_PRL, bender2020fluctuations}, and Anderson localization~\cite{van1999multiple, lagendijk2009fifty, hildebrand2014observation}. They also play a fundamental role in wavefront shaping by limiting and modifying the coherent control of waves in complex scattering media~\cite{2012_Choi_Nat._Photonics, 2014_Popoff_PRL, sarma2016control, 2017_Wade_Nat._Phys., 2019_Yilmaz_Nat._Photonics, bender2020fluctuations, shaughnessy2024multiregion}. An outstanding question is whether long-range mesoscopic correlations influence the second-order statistics of the enhancement factor. 

Our experiments and numerical simulations address this question by revealing correlation-induced giant fluctuations that exceed the predictions of the Laguerre--Wishart (LW) random-matrix model, thereby identifying correlation-driven deviations from the LW baseline. From a fundamental perspective, our results demonstrate that the enhancement factor is not merely a measure of focusing performance, but also a statistical observable whose fluctuations encode long-range mesoscopic correlations in wavefront-shaping experiments. This establishes enhancement-factor fluctuations as a simple and experimentally accessible probe of long-range mesoscopic correlations. From an applied perspective, the finite-size LW framework provides accurate parameter-free predictions for the enhancement factor and its fluctuations when the influence of long-range mesoscopic correlations is negligible, while providing a correlation-free baseline for identifying excess fluctuations when such correlations become significant. This framework, therefore, enables more reliable performance estimates for applications in biomedical imaging, optical metrology, optical trapping, and optical communications through scattering media.

The remainder of this paper is organized as follows. In the Results section, we first present the experimental enhancement-factor statistics and compare them with the exact finite-size Laguerre--Wishart predictions. We then introduce the random-matrix framework based on the Laguerre--Wishart ensemble and provide exact predictions for the enhancement-factor statistics. Next, we investigate the role of long-range mesoscopic correlations by combining numerical simulations, analytical predictions, and experimental measurements, demonstrating that these correlations give rise to giant fluctuations in the enhancement factor beyond the Laguerre--Wishart baseline. In the Discussion section, we place our findings in a broader context and summarize the main results. Finally, the Methods section provides details of the experimental transmission-matrix measurements and wave-propagation simulations.

\begin{figure*}[th]
\centering
\includegraphics[width=\linewidth]{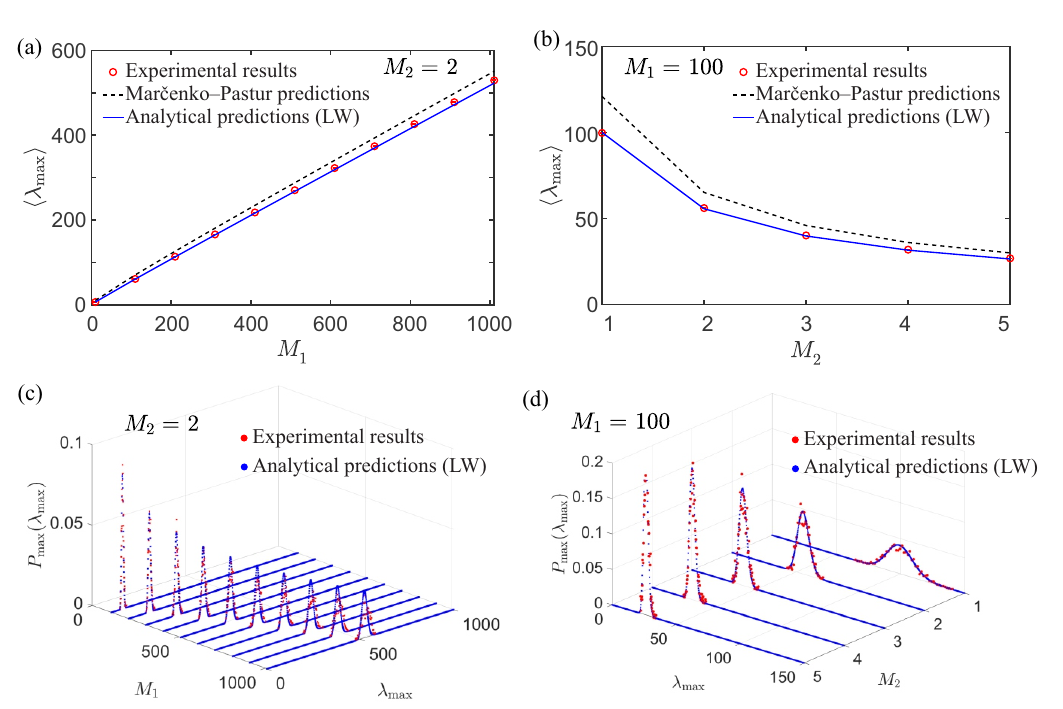}
\caption{The mean and probability density of the enhancement factor $\lambda_{\max}$ are shown. The mean enhancement $\langle\lambda_{\max}\rangle$ is plotted versus the number of input channels $M_{1}$ at fixed $M_{2}=2$ (a) and versus the number of output channels $M_{2}$ at fixed $M_{1}=100$ (b). The predictions of the finite-size Laguerre--Wishart (LW) ensemble (solid blue), obtained from the exact analytical expression, are in quantitative agreement with the experiments (red open circles). In contrast, the Mar\v{c}enko--Pastur prediction (dashed black) lies well outside the experimental uncertainty in panels (a,b). The error bars, computed from three independent experimental transmission-matrix measurements, are smaller than the open-circle symbols. With no fitting parameter, the predicted probability densities $P(\lambda_{\max}) = Q'_{\max}(\lambda_{\max})$ from Eq. (\ref{Qmax}) likewise match quite well the experimental distributions for small $M_1$ in (c) and for all values of $M_2$ for $M_1 = 100$ in~(d).}
\label{figure2}
\end{figure*}

\section{Results}
\subsection{Transmission-matrix measurements and enhancement statistics}

The transmission matrix of a scattering medium provides the information required to modulate the incident wavefront with a spatial light modulator (SLM) and focus light onto single or multiple target points, as illustrated in Fig.~\ref{figure1}: an unmodulated beam produces a random speckle pattern [Fig.~\ref{figure1}(a)], whereas a shaped input beam enables simultaneous focusing at multiple target points [Fig.~\ref{figure1}(b)].

Experimentally, we measure the transmission matrix $t$ of a scattering sample using phase-shifting interferometry~\cite{yilmaz2021customizing}, see Supplementary Information for a sketch of the experimental setup. The scattering sample consists of a densely packed zinc oxide (ZnO) nanoparticle layer on a cover slip, with a thickness of approximately 10 \textmu m, much larger than the transport mean free path $l_t = 1$ \textmu m, ensuring that light transport within the ZnO layer is diffusive. In our measured $t_{mn}$, the input channel $n$ corresponds to a $9\times 9$ SLM segment ($\approx$ one Fourier component on the front surface). The output channel $m$ corresponds to a position on the back surface, i.e., $t_{mn}$ maps the SLM field to the outgoing field at the back surface.

Starting from the experimental transmission matrix $t_{mn}$, we generate an ensemble of focusing operators $A^\dagger A$, with $A$ denoting a randomly chosen sub-transmission matrix of size $M_2 \times M_1$. We work in the regime $M_1 \gg M_2$, with $M_1 \in [100,1000] \cap \mathbb{Z}$ and $M_2 \in [1,5] \cap \mathbb{Z}$. Each sub-matrix $A$ is formed by randomly selecting $M_1$ input segments on the SLM and $M_2$ output target focusing positions $m$ in transmission. For each sub-matrix size $M_2 \times M_1$, we compute its eigenvalues via $A^\dagger A V_\alpha = \lambda_\alpha V_\alpha$. We then determine the probability density $P_{\rm max}(\lambda_{\max})$, the mean $\langle \lambda_{\max}\rangle$, and the standard deviation $\Delta\lambda_{\max}$ of the largest eigenvalue of the associated focusing operator, thereby obtaining the experimental statistical distribution of the enhancement factor for light focusing through the scattering sample.

In Fig.~\ref{figure2}, we present the experimental and theoretical largest eigenvalues \(\lambda_{\rm max}\) (i.e., enhancement values \(\eta\)) and their probability densities \(P_{\rm max}(\lambda_{\rm max})\) for various numbers of input \((M_1)\) and output \((M_2)\) channels. The scaling of \(\lambda_{\rm max}\) with \(M_1\) [Fig.~\ref{figure2}(a)] and with \(M_2\) [Fig.~\ref{figure2}(b)] predicted by the LW random-matrix-based analytical calculation is in excellent agreement with the experimental values, whereas the Mar\v{c}enko--Pastur prediction (\ref{eq:1}) shows a clear deviation. This deviation arises because the Mar\v{c}enko--Pastur law is exact only in the asymptotic limit of large \(M_1\) and \(M_2\), while the LW random-matrix-based analytical calculation remains accurate for finite \(M_1\) and \(M_2\). Moreover, the LW-based random matrix theory predicts not only the mean of \(\lambda_{\rm max}\) but also its full distribution \(P_{\rm max}(\lambda_{\rm max})\), as shown in Figs.~\ref{figure2}(c,d). The theoretical prediction of $P_{\max}(\lambda_{\max})$ agrees well with the experimental distribution at small $M_1$. As $M_1$ increases, however, the experimentally observed distribution becomes progressively broader than the Laguerre--Wishart prediction (Fig.~\ref{figure2}(c)). Within the correlation-free Laguerre--Wishart framework, this systematic excess broadening is the first indication that long-range mesoscopic correlations contribute to enhancement-factor fluctuations. This interpretation is consistent with the well-established increase of long-range mesoscopic correlations with the increasing number of controlled input channels $M_1$~\cite{2008_Vellekoop_PRL, 2013_Stone_PRL, 2014_Popoff_PRL, 2017_Wade_Nat._Phys., shaughnessy2024multiregion}. For $M_1=100$, the experimental and theoretical probability densities remain in close agreement for all investigated values of $M_2\in[1,5]\cap\mathbb{Z}$ (Fig.~2(d)).

\subsection{Random-matrix framework: the Laguerre–Wishart ensemble}

\begin{figure*}[th]
\centering
\includegraphics[width=\linewidth]{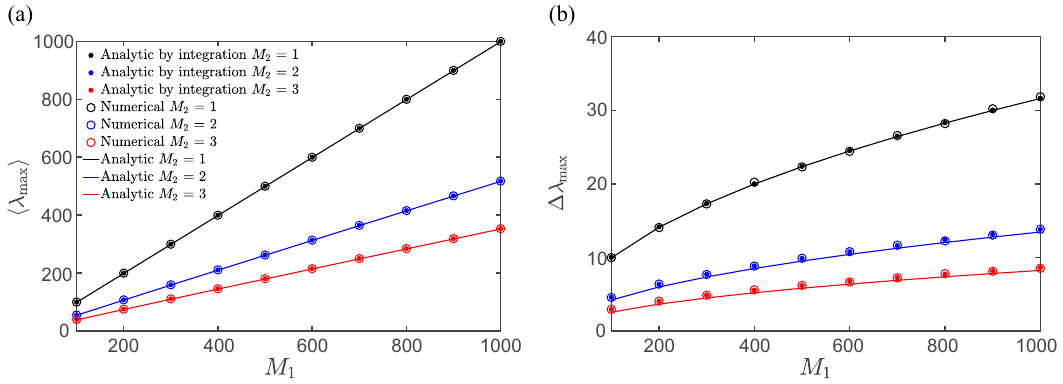}
\caption{Mean and standard deviation of the largest eigenvalue in the Laguerre--Wishart ensemble. Panels (a) and (b) show respectively the mean and the standard deviation of the largest eigenvalue, $\lambda_{\max}$ (i.e., the enhancement factor), as a function of the number of input channels $M_1$, for different numbers of target channels $M_2 = 1, 2, 3$ (indicated by color). Open circles denote numerical results obtained from Laguerre–Wishart (LW) matrices, while solid circles correspond to the exact analytical predictions derived from the cumulative distribution function of $\lambda_{\max}$ [Eq. (\ref{Qmax})]. Solid lines show the asymptotic large-$M_1$ results [Eqs. (\ref{large_nu_main}) and (\ref{std_dev})], demonstrating excellent agreement between numerical, exact, and asymptotic results.}
\label{figure3}
\end{figure*}
    
Having presented the experimental results for the mean enhancement factor and its probability distributions, together with their comparison to the analytical predictions, we now detail the random-matrix framework from which these predictions are derived. Our approach is based on an analytic treatment of the spectral properties of the focusing operator within a random-matrix framework. To this end, we model the entries of the $M_2 \times M_1$ sub-transmission matrix $A$ as independent complex Gaussian variables. This is the standard statistical description of open scattering systems that support a large number of channels, in which the coherent superposition of many scattering paths gives rise to fully developed speckle and approximately Gaussian field statistics~\cite{goodman2015statistical, fyodorov2023intensity, kohnes2026universality, fyodorov2026random}. The resulting Gaussian model provides the simplest parameter-free description of enhancement-factor statistics in the absence of long-range mesoscopic correlations, serving as a quantitative reference for identifying correlation-induced excess fluctuations. Under this assumption, the focusing operator $A^\dagger A$ belongs to the complex anti--LW ensemble~\cite{beenakker1997random}, where ``anti'' indicates that the underlying matrix $A$ is rectangular with more columns than rows ($M_1 > M_2$)~\cite{yu2002anti,janik2003wishart,vivo2007large}. Consequently, $A^\dagger A$ is an $M_1 \times M_1$ random matrix whose spectrum consists of exactly $M_2$ positive eigenvalues $\lambda_1,\dots,\lambda_{M_2}$, while the remaining $M_1 - M_2$ eigenvalues are identically zero.

These nonzero eigenvalues encode the strength of the transmission channels. Since the matrix $A$ is random, so are these nonzero eigenvalues, which
thus fluctuate from one sample of $A$ to another. However, it turns out that 
their joint (i.e., multivariate) statistics are known exactly. Indeed, their joint probability density function (JPDF) reads \cite{yu2002anti,janik2003wishart}
\begin{equation} \label{JPDF}
P(\lambda_1,\dots,\lambda_{M_2}) = \frac{1}{Z} \prod_{i=1}^{M_2} \lambda_i^{\nu}\, e^{-M_2 \lambda_i} \prod_{i<j} (\lambda_i - \lambda_j)^2 ,
\end{equation}
where $\nu = M_1 - M_2 >0$ and $Z$ is a normalization factor that can be computed explicitly. 
%$Z = M_2^{-M_1 M_2} \prod_{k=0}^{M_2-1} [(k+1)!(k+\nu)!]$ and $\nu = M_1 - M_2 > 0$ \cite{ref}. 
From this JPDF~(\ref{JPDF}), the average density of eigenvalues, defined as $\rho(\lambda) = \sum_{i=1}^{M_1} \langle \delta(\lambda-\lambda_i)\rangle$ where $\delta$ is a Dirac delta-function and $\langle \cdots \rangle$ denotes an average over the JPDF in (\ref{JPDF}), can be computed explicitly. It is simply obtained by integrating the JPDF (\ref{JPDF}) over $\lambda_2, \cdots, \lambda_{M_2}$ while keeping $\lambda_1 \equiv \lambda$ fixed. This multiple integral can be carried out explicitly and 
it reads $\rho(\lambda) = \nu \delta(\lambda) + \tilde \rho(\lambda)$ where the density of nonzero eigenvalues is given by (see e.g. \cite{vivo2008wishart})
\begin{equation} \label{density}
\tilde \rho(\lambda) = M_2^{\nu+1 }\lambda^\nu e^{-M_2 \lambda} \sum_{k=0}^{M_2-1} \frac{k!}{(k+\nu)!} [L_k^{(\nu)}(M_2 \lambda)]^2 \;,
\end{equation}
where $L_k^{(\nu)}(\lambda)$ is the generalized Laguerre polynomial of index $\nu$ and degree $k$. 

Of particular interest is the statistics of the largest eigenvalue 
$\lambda_{\max} = \max \{\lambda_1, \dots, \lambda_{M_2}\}$, 
which in the context of wavefront shaping corresponds directly to the enhancement factor~\cite{2017_Wade_Nat._Phys.}. Rather than computing its probability density function $P_{\max}(x)$ directly, 
it is often more convenient to consider the cumulative distribution function 
$Q_{\max}(x) = \mathrm{Prob}(\lambda_{\max} \le x) = \int_0^x P_{\max}(x')\,dx'$ \cite{majumdar2024statistics}. Indeed, the event $\lambda_{\max} \le x$ is equivalent to all eigenvalues satisfying 
$\lambda_i \le x$ for $i = 1, \dots, M_2$. Therefore, $Q_{\max}(x)$ can be obtained by integrating 
the joint probability density function in (\ref{JPDF}) over the hypercube $[0,x]^{M_2}$. 
This multiple integral can be evaluated explicitly using standard techniques from 
random matrix theory and determinantal processes, yielding (see also \cite{dighe2003analysis,forrester2019recursion})
\begin{equation} \label{Qmax}
Q_{\max}(x) = A_{M_1,M_2}\det_{1\leq i,j \leq M_2} \Big[ \gamma(i+j-1+\nu, M_2 x) \Big] \;,
\end{equation}
where $\gamma(s,z) = \int_0^z t^{s-1} e^{-t} \, dt$ is 
the lower incomplete gamma function (which in our case can be simply expressed in terms of exponentials and polynomials) and $A_{M_1, M_2}=M_2!/(\prod_{k=0}^{M_2-1} (k+1)! \, (k+\nu)!)$ is a normalization constant (independent of $x$). As $x \to 0$ it behaves as $Q_{\max}(x) \sim x^{M_1 M_2}$ while $1-Q_{\max}(x) \sim x^{M_1 + M_2-2}\, e^{-M_2 x}$ as $x \to \infty$.

\begin{figure*}[th]
\centering
\includegraphics[width=\linewidth]{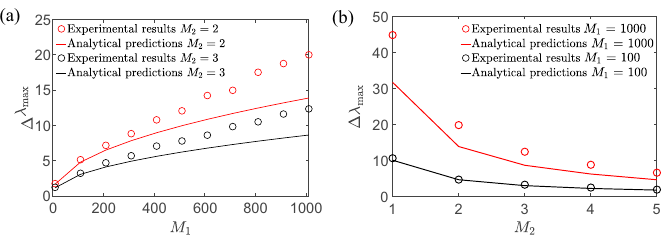}
\caption{Experimental enhancement-factor fluctuations and Laguerre--Wishart predictions. Fluctuations of the enhancement factor, quantified by the standard deviation $\Delta\lambda_{\max}\equiv\sqrt{\langle\lambda_{\max}^2\rangle-\langle\lambda_{\max}\rangle^2}$, are shown versus the number of input channels $M_{1}$ at fixed $M_2 = 2$ and $M_2 = 3$ (a) and versus the number of output channels $M_{2}$ at fixed $M_{1}=100$ and $M_{1}=1000$ (b). The LW-ensemble theory (solid lines) agrees well with the experiments (open circles) up to $M_{1}\approx 100$ in (a) and for all $M_{2}$ at fixed $M_1 = 100$ in (b). For $M_{1}>100$, the experiments exhibit larger fluctuations than predicted (a,b), indicating systematic deviations from the correlation-free LW model.}
\label{figure4}
\end{figure*}

This $M_2 \times M_2$ determinant in Eq. (\ref{Qmax}) is straightforward to evaluate numerically (for moderate values of $M_2$), allowing a direct computation of the mean $\langle \lambda_{\max} \rangle$ 
and fluctuations $\Delta \lambda_{\max}$. This exact characterization (\ref{Qmax}) is 
particularly useful in wavefront shaping experiments, where $M_2$ is typically small. As discussed in the previous subsection, Fig.~\ref{figure2} revisits the behavior of the largest eigenvalue, but now emphasizes the exact finite-size analytical predictions. In Fig.~\ref{figure2}(a,b), we use Eq.~(\ref{Qmax}) to compute the theoretical curves for $\langle \lambda_{\max} \rangle$ and compare them with the experimental values, finding excellent agreement for all investigated $M_1$ and $M_2$. At the same time, these panels clearly reveal a deviation from the MP prediction (\ref{eq:1}), which is expected since the MP law strictly applies in the limit where both $M_1$ and $M_2$ are large, whereas here $2 \leq M_2 \leq 5$. In Fig.~\ref{figure2}(c,d), we further use $P_{\max}(x) = dQ_{\max}(x)/dx$ with $Q_{\max}(x)$ given in Eq.~(\ref{Qmax}) to generate the analytical curves for the full distribution of $\lambda_{\max}$ and compare them with the experimentally obtained histograms, finding excellent agreement for small $M_1$.

In wavefront shaping experiments, one is often interested in the limit where $M_1 \gg M_2$. In this limit, one can show that the complex LW ensemble in (\ref{JPDF}) converges, up to a rescaling of the eigenvalues by $M_1$ after centering around $M_1/M_2$ and rescaling by $M_2/\sqrt{2 M_1}$, to the so-called Gaussian Unitary Ensemble (GUE) that describes Hermitian matrices with independent (complex) Gaussian entries. Consequently for large $M_1 \gg  M_ 2$ one finds
\begin{eqnarray} \label{large_nu_main}
\langle \lambda_{\max} \rangle \approx \frac{1}{M_2}\left( M_1 +\sqrt{2 \,M_1} \, \langle \mu_{\max}\rangle \right) \;,
\end{eqnarray}
where $\langle \mu_{\max}\rangle$, which only depends on $M_2$ and not on $M_1$, denotes the average value of the largest eigenvalue of a GUE matrix of size $M_2 \times M_2$. For example, one finds $\langle \mu_{\rm max} \rangle = \{0, \sqrt{2/\pi}, 27/(8 \sqrt{2\pi})\}$ for $M_2 = \{1,2,3\}$  respectively (see Supplementary Information for further details). In the limit $M_2 \gg 1$ one has $\langle \mu_{\max}\rangle \approx \sqrt{2 M_2}$ and this formula (\ref{large_nu_main}) coincides at leading order with the Mar\v{c}enko-Pastur prediction given in Eq. (1), in the limit $M_1/M_2 \gg 1$. However, for finite $M_2$, the average value $\langle \mu_{\max} \rangle$ differs from $\sqrt{2 M_2}$ and therefore the large $M_1$ estimate in (\ref{large_nu_main}) differs from the Mar\v{c}enko-Pastur prediction. This formula (\ref{large_nu_main}) ultimately enables a more accurate description of experimental data. 

Similarly, by exploiting the convergence of the complex LW ensemble to the GUE in the limit $M_1 \gg M_2$, one can obtain the behavior of the standard deviation $\Delta \lambda_{\max}$ in this limit. One finds indeed
\begin{equation} \label{std_dev}
\Delta \lambda_{\max} \approx {\frac{\sqrt{2 M_1}}{M_2}} \Delta \mu_{\max} \;,
\end{equation}
where $\Delta \mu_{\max}$, which depends only on $M_2$, is the standard deviation of the largest eigenvalue of a GUE matrix of size $M_2 \times M_2$. It can be computed explicitly for the first few values of $M_2$ (see Supplementary Information for further details). These asymptotic behaviors for the mean (\ref{large_nu_main}) and the standard deviation (\ref{std_dev}) are compared to numerical and exact calculations in Fig.~\ref{figure3}, showing good agreement.

We conclude this subsection by noting that in the opposite limit, when both $M_1$ and $M_2$ are large with a fixed ratio, the eigenvalue density converges to the Mar\v{c}enko--Pastur law, supported on $[x_-, x_+]$ with $x_\pm = (\sqrt{M_1/M_2} \pm 1)^2$. In this limit, $\lambda_{\max}$ concentrates near $x_+$, with fluctuations of order $\Delta\lambda_{\max} \sim M_1^{1/2} M_2^{-7/6}$, with $\Delta \mu_{\max} \sim M_2^{-1/6}$, governed by the celebrated Tracy--Widom (TW) distribution~\cite{tracy1994level,johansson2000shape,johnstone2001distribution} (for a review see~\cite{majumdar2014top}), recovering the approximation used in Eq.~(\ref{eq:1}). The familiar Tracy--Widom scaling $\Delta\lambda_{\max}\sim M^{-2/3}$ is recovered as the special case $M_1=M_2=M$. The prefactor governing the scaling of $\Delta\lambda_{\max}$ can be expressed in terms of the standard deviation of the Tracy–Widom distribution. Although no elementary closed-form expression is known for this constant, its value can be accurately evaluated numerically. This formula~(\ref{Qmax}) thus provides both the exact finite-$M_2$ statistics of $\lambda_{\max}$ and the asymptotic large-$M_1, M_2$ behavior, bridging theoretical insight and experimental relevance. Extreme eigenvalue statistics in the TW regime have been observed experimentally in other physically distinct systems in the large-size limit, such as interface growth in nematic liquid crystals~\cite{takeuchi2011growing}, coupled lasers~\cite{fridman2012measuring}, and dissipative many-particle systems~\cite{makey2020universality}. In contrast, our experiments probe the largest eigenvalue of finite-size LW matrices in a wavefront-shaping setting, where the distribution generically deviates from TW statistics.

\subsection{The role of long-range mesoscopic correlations in enhancement-factor fluctuations}

\begin{figure*}[tb]
\centering
\includegraphics[width=\linewidth]{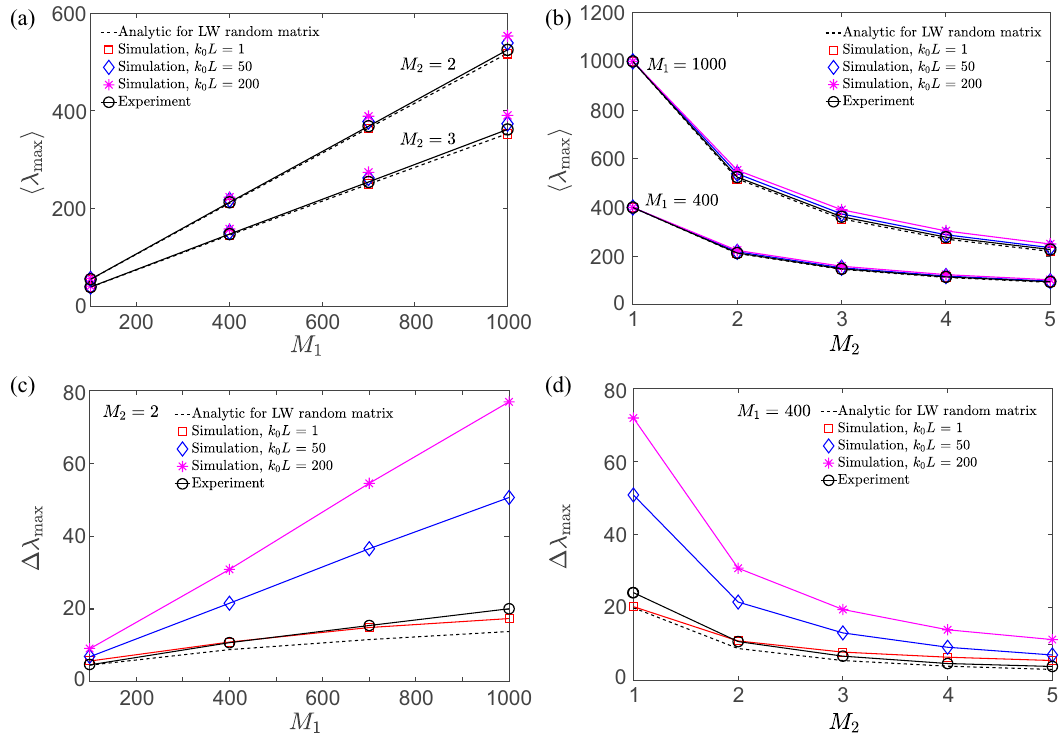}
\caption{Mean enhancement and giant fluctuations in the presence of long-range mesoscopic correlations. The mean enhancement $\langle\lambda_{\max}\rangle$ versus the number of input channels $M_{1}$ at fixed $M_{2}=2$ and $M_{2}=3$ (a) and versus the number of output channels $M_{2}$ at fixed $M_{1}=400$ and $M_{1}=1000$ are shown for the analytic LW-ensemble theory (dashed black line), experiments (black open circles), and numerical simulations (red open squares for $k_0L = 1$, blue open diamonds for $k_0L = 50$, and magenta asterisks for $k_0L = 200$, all with normalized transport mean free path $k_0 l_t = 20.6$). All results agree across $M_{1}$ and $M_{2}$, consistent with the mean enhancement being insensitive to long-range mesoscopic correlations. Fluctuations, quantified by the standard deviation $\Delta\lambda_{\max}\equiv\sqrt{\langle\lambda_{\max}^{2}\rangle-\langle\lambda_{\max}\rangle^{2}}$ versus the number of input channels $M_{1}$ at fixed $M_{2}=2$ (c), and versus the number of output channels $M_{2}$ at fixed $M_{1}=400$ (d), exhibit a slight experimental excess and a pronounced excess in simulations, which grows with slab thickness. These giant fluctuations in thick slabs reveal long-range mesoscopic correlations beyond the correlation-free LW model.}
\label{figure5}
\end{figure*}

Our analytical theory provides the full probability distribution of the largest eigenvalue and thus predicts its fluctuations through the standard deviation \(\Delta \lambda_{\max}\). In Figs.~\ref{figure4}(a,b), we plot the experimentally measured \(\Delta \lambda_{\max}\) as functions of \(M_1\) and \(M_2\), together with the correlation-free LW-based predictions. For small input-channel numbers, \(M_1 \lesssim 100\), the theory reproduces the experimental fluctuations. For larger \(M_1\), however, the theory increasingly underestimates the fluctuations, consistent with the broader experimental \(P_{\rm max}(\lambda_{\rm max})\) compared to the LW prediction (Fig.~\ref{figure2}(c)).

The LW-based framework is built on correlation-free random matrices. In contrast, it is well established that the transmission matrix of a disordered medium exhibits long-range mesoscopic correlations~\cite{2007_Akkermans, berkovits1994correlations, van1999multiple, 1987_Cwilich, 1988_Stone, 1988_Mello_Akkermans_Shapiro, 1989_Shapiro, 1990_Genack, 2013_Muskens}. The systematic excess of the experimental fluctuations over the LW prediction in Figs.~\ref{figure4}(a,b) therefore naturally raises the question of whether the observed fluctuations of the largest eigenvalue (i.e., the enhancement factor) in wavefront-shaping focusing experiments originate from such long-range correlations.

To address this question and systematically assess the role of long-range correlations—which are known to increase with sample thickness~\cite{2007_Akkermans, berkovits1994correlations, van1999multiple, 2017_Wade_Nat._Phys.}—we performed numerical simulations for slabs of varying thicknesses. Specifically, we computed the field transmission matrix \(t\) of two-dimensional (2D) scattering slabs with normalized thicknesses \(k_0 L = 1, 50, 200\) and normalized transport mean free path \(k_0 l_t = 20.6\) using a recursive Green's function algorithm~\cite{1985_MacKinnon, 1991_Stone}, where $k_0 = 2\pi/\lambda_0$ denotes the wavenumber corresponding to the wavelength $\lambda_0$.

For each slab thickness, we systematically obtained the mean largest eigenvalue $\langle \lambda_{\max} \rangle$—averaged over different focusing target positions—and its standard deviation $\Delta \lambda_{\max}$ for sub-transmission matrices of various sizes $M_2 \times M_1$ (see Supplementary Information).

In Fig.~\ref{figure5}, we show the numerical, analytical, and experimental results for both the mean and the standard deviation of the largest eigenvalue. As seen in Figs.~\ref{figure5}(a,b), the numerical mean largest eigenvalue $\langle \lambda_{\rm max} \rangle$ (i.e., the enhancement factor $\eta$) scales with the number of input channels $M_1$ and the number of output channels $M_2$ in excellent agreement with the LW-random-matrix-based analytical results and with the experimental results, for all slab thicknesses. This behavior is consistent with the mean enhancement factor being insensitive to long-range mesoscopic correlations~\cite{2008_Vellekoop_PRL, 2017_Wade_Nat._Phys.} and, consequently, to the slab thickness.

The situation is different for the fluctuations. In Figs.~\ref{figure5}(c,d), we observe a clear increase in the standard deviation $\Delta \lambda_{\rm max}$ of the largest eigenvalues as the slab becomes thicker in the simulations, which is a clear signature of long-range mesoscopic correlations governing these fluctuations. Moreover, our simulation results show that the observed fluctuations can be several times larger than the correlation-free LW prediction, thus indicating that long-range mesoscopic correlations can give rise to giant fluctuations of the enhancement factor, far beyond the finite-size random-matrix baseline.

Remarkably, even for extremely thin slabs ($L \ll l_t$), the fluctuations already start to deviate from the correlation-free prediction as $M_1$ increases (Fig.~\ref{figure5}(c)), and the experimental data show the same trend. One might expect even larger deviations in the experiment, since the sample is much thicker than a transport mean free path ($L \gg l_t$). However, the observable strength of long-range correlations decreases with the input-control ratio $M_1/g$, where $g$ is the dimensionless conductance or number of open channels~\cite{2014_Popoff_PRL, 2013_Stone_PRL, 2017_Wade_Nat._Phys.}. Because our experimental slab is three-dimensional (3D), $g$ is much larger than in the 2D simulations, and therefore the effective input-control ratio is substantially smaller. This reduced control weakens the impact of long-range correlations on $\Delta\lambda_{\rm max}$ in the experiment, explaining the quantitative differences while preserving the qualitative agreement.

Traditionally, long-range correlations are revealed through the variance of the eigenvalue density of $t^\dagger t$ or through shifts in $\langle \lambda_{\rm max} \rangle$, but this requires measuring large transmission matrices with $M_1 \gtrsim 2000$ inputs and $M_2 \gtrsim 3000$ outputs through very thick samples ($\approx 50 l_t$)~\cite{2017_Wade_Nat._Phys.}. In contrast, we show here that long-range correlations can be detected from much smaller datasets—using submatrices with only $M_1 \lesssim 200$ and even a single output channel ($M_2 = 1$)—and through samples as thin as $\approx 10 l_t$. This is possible because the LW-based analytical framework provides exact predictions in the absence of correlations, allowing systematic excess fluctuations beyond the correlation-free baseline to serve as a sensitive signature of long-range mesoscopic correlations.

\section{Discussion}

In conclusion, our work establishes the statistical foundations of the enhancement factor, a key measure of coherent control of light through complex scattering media with implications for biomedical imaging, optical metrology, secure communication, optical trapping, and quantum photonics. By combining experiments, numerical simulations, and exact random-matrix results, we determine the full probability distribution of the enhancement factor for multi-point focusing. For finite numbers of input and output channels, and within the Laguerre--Wishart (LW) random-matrix framework (i.e., in the absence of long-range mesoscopic correlations), the theory quantitatively captures both the mean and the fluctuations, going beyond the Mar\v{c}enko--Pastur prediction for the mean and the Tracy--Widom law for fluctuations in the asymptotic large-channel limit. These results provide, for the first time, a comprehensive statistical description directly relevant to experimentally accessible finite-channel regimes.

Beyond applying the Laguerre--Wishart framework to enhancement-factor statistics in wavefront shaping, we demonstrate that strongly scattering media exhibit giant fluctuations arising from long-range mesoscopic correlations. While probing long-range mesoscopic correlations through the mean largest eigenvalue would normally require nearly complete transmission matrices containing $10^{4}$--$10^{6}$ channels~\cite{2012_Choi_Nat._Photonics, 2013_Park, 2013_Stone_PRL, 2014_Popoff_PRL,2017_Wade_Nat._Phys.}, our results show that enhancement-factor fluctuations already provide a sensitive probe when only $M_1 \lesssim 200$ input and $M_2 = 1$ output channels are controlled. This establishes a direct connection between wavefront-shaping experiments and mesoscopic wave transport, revealing the limitations of correlation-free random-matrix predictions in the presence of long-range mesoscopic correlations.

A natural question is whether these correlation-induced deviations can instead be described within the framework of correlated Wishart ensembles~\cite{simon2004eigenvalue, burda2005spectral, simon2006capacity, zanella2009marginal}. Such models introduce correlations through the second-order covariance structure of the Gaussian transmission matrix, $\left\langle t_{j_1i_1}^{*}t_{i_2j_2}\right\rangle$, which characterizes pairwise correlations between transmission coefficients, whereas the long-range mesoscopic correlations considered here arise from the coherent interference of multiple scattering paths and are characterized by higher-order four-field connected correlation functions (i.e., fourth-order cumulants), $\left\langle t_{j_1i_1}^{*}t_{i_2j_2}t_{j_3i_3}^{*}t_{i_4j_4}\right\rangle_c$~\cite{2007_Akkermans, berkovits1994correlations, van1999multiple}, rather than solely by second-order covariances. Consequently, these physical correlations cannot, in general, be represented solely through a prescribed covariance matrix for Gaussian transmission coefficients. Existing theories either explicitly incorporate long-range mesoscopic correlations into random-matrix models while describing only the global eigenvalue density~\cite{2013_Stone_PRL}, or accurately reproduce the global spectrum without explicitly encoding these correlations, as in radiant-field theory~\cite{gaspard2025radiant,gaspard2025transmission}. Establishing a predictive mapping between mesoscopic multiple-scattering theory and correlated random-matrix theory—and ultimately an accurate theory for the extreme eigenvalue statistics in the presence of long-range mesoscopic correlations—therefore remains an important open problem.

Although demonstrated here for optical wavefront shaping, the underlying mechanism relies only on multiple scattering and finite-size statistics, suggesting that the fluctuation physics uncovered in this work should extend naturally to acoustic, elastic, microwave, and matter-wave systems. More broadly, our results establish the enhancement factor not merely as a measure of focusing performance, but as a statistical observable that provides a simple and experimentally accessible probe of long-range mesoscopic correlations in complex scattering media.

\begin{figure*}[t]
	\centering
	\includegraphics[width=\linewidth]{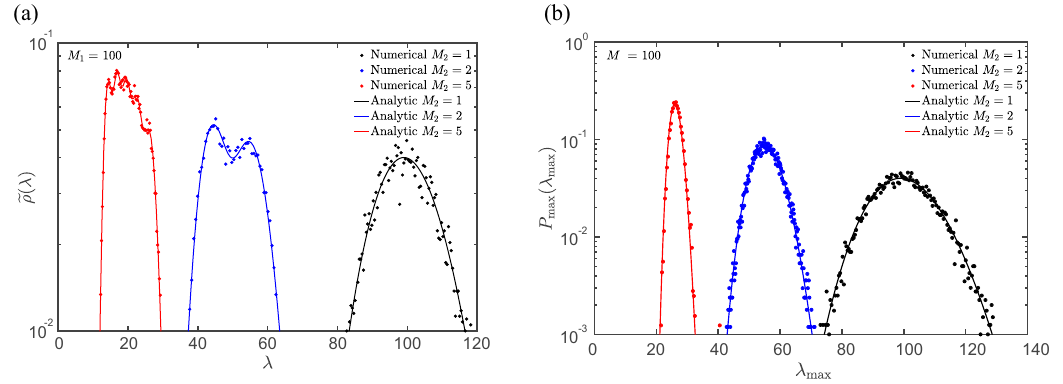}
	\caption{(a) Eigenvalue distributions $\widetilde{\rho}(\lambda)$ and (b) largest-eigenvalue distributions $P_{\rm max}(\lambda_{\rm max})$ for $M_1 = 100$ and $M_2 = 1, 2, 5$. The LW–random-matrix-based predictions and the numerical Wishart-matrix results coincide remarkably well for both $\widetilde{\rho}(\lambda)$ given in (\ref{CD}) and $P_{\rm max}(\lambda_{\rm max}) = Q'_{\max}(\lambda_{\max})$ given in (\ref{Qmax_expl1}).}
	\label{figureS1}
\end{figure*}

\section{Methods}
\subsection{Experimental setup and transmission-matrix measurements}
We measure the field transmission matrix of a $\sim$10 \textmu m-thick layer of densely packed ZnO nanoparticles (with average diameter $\sim$200 nm) deposited on a 170 \textmu m-thick cover slip (average transmittance $\sim$0.2 at $\lambda=532$ nm) using a Mach–Zehnder interferometer (shown in Fig.~\ref{figureS3}). A linearly polarized 532-nm laser beam (Coherent, Compass 215M-50 SL) is expanded, collimated, and split into two arms; in the sample arm, the beam reflects off a spatial light modulator (Hamamatsu X10468-01) and is imaged onto the pupil of a microscope objective ${\rm MO}_1$ (Nikon 100$\times$, ${\rm NA}_{\rm in}=0.95$) by lenses $L_1$ ($f_1=100$ mm) and $L_2$ ($f_2=250$ mm), with an iris blocking higher diffraction orders, such that each SLM segment corresponds approximately to a plane wave incident on the sample surface at a specific angle. The transmitted light is collected by an oil-immersion microscope objective ${\rm MO}_2$ (Edmund 100$\times$, ${\rm NA}_{\rm out}=1.25$), collimated by $L_5$ ($f_5=200$ mm) and $L_6$ ($f_6=150$ mm), and interfered with the reference beam; the interference pattern is recorded at the Fourier plane of the sample output by a CCD camera (Allied Vision Manta G-031B), after a single linear polarization selection.

The transmission matrix is measured in the Hadamard basis using four-phase-shift interferometry. A circular aperture on the SLM covers 2415 segments, from which $32\times32=1024$ central segments (each $9\times9$ pixels) are used as input channels, while a high-frequency grating outside this region diffracts light away. Four global phase shifts, equally spaced between $0$ and $2\pi$, are applied to the Hadamard patterns to extract the relative phase with the reference beam; blocking the reference allows for intensity measurements of all inputs. Combining these yields the complex transmitted field, which is detected on a $64\times64=4096$-pixel CCD region, giving a transmission matrix $t_{m'n'}$ of size $4096\times1024$. We then apply a Hadamard transform as $t_{m'n} = \sum_{n'=1}^{N} t_{m'n'}H_{n'n}$ to convert the transmission matrix into the canonical (SLM pixel) basis $n$. Here $H_{n'n}$ represents the unitary Hadamard transform matrix. In our experimental setup, each output index $m'$ corresponds to a CCD camera pixel, representing a single Fourier component on the sample back surface. Since the relative phase between each output channel $m'$ is known, a two-dimensional Fourier transform can be applied to each column of the transmission matrix $t_{m'n}$ to obtain the transmission matrix $t_{mn}$, where $m$ denotes the position on the back surface of the sample, mapping the light field on the SLM to the outgoing field at the back surface of the sample.

After obtaining the experimental transmission matrix $t_{mn}$, where $n$ denotes the SLM segment index and $m$ the position index on the sample back surface, for each choice of $M_2$, we normalize the transmission-matrix elements such that ${\rm Var}(t_{mn}) = 1/M_2$. We then construct an ensemble of randomly selected sub-transmission matrices $A$ of size $M_2 \times M_1$. We focus on the regime $M_1 \gg M_2$, with $M_1 \in [100,1000] \cap \mathbb{Z}$ and $M_2 \in [1,5] \cap \mathbb{Z}$. Each sub-matrix $A$ is formed by randomly choosing $M_1$ input segments on the SLM and $M_2$ output target positions in transmission. Since the mean transmitted intensity profile on the back surface of the sample is not spatially uniform due to diffusion, we restrict the rows of $A$ to the central region of the diffuse spot, where the profile is approximately uniform, and select the target positions accordingly.

\subsection{Wave-propagation simulations}

In our simulations, we model wave propagation through two-dimensional (2D) diffusive slabs of size $k_0 W \times k_0 L$, where $W$ and $L$ denote the slab width and length, respectively. The sample is discretized on a 2D square grid with grid spacing $(\lambda_0/2\pi)\times(\lambda_0/2\pi)$. The dielectric constant at each grid point is defined as
\begin{equation}
    \epsilon(\mathbf{r}) = n_0^2 + \delta\epsilon(\mathbf{r}),
\end{equation}
where $n_0$ is the average refractive index of the medium, and $\delta\epsilon(\mathbf{r})$ is a uniformly distributed random number in the range $[-\Sigma,\Sigma]$. The slab is embedded between two homogeneous media of refractive indices $n_1$ and $n_2$, and periodic boundary conditions are imposed along the transverse direction. The field transmission matrix $t_{mn}$ at wavelength $\lambda_0$ is computed by solving the scalar wave equation
\begin{equation}
    \big[\nabla^2 + k_0^2 \epsilon(\mathbf{r})\big]\psi(\mathbf{r}) = 0,
\end{equation}
with $k_0 = 2\pi/\lambda_0$ the wavenumber corresponding to the wavelength $\lambda_0$, using the recursive Green’s function method~\cite{1985_MacKinnon,1991_Stone}.

Starting from the computed transmission matrix $t_{mn}$, where $n$ denotes the index of incident Fourier components at the input and $m$ the spatial position at the output surface of the scattering slab, we construct an ensemble of focusing operators $A^\dagger A$. Here, $A$ is a randomly selected sub-transmission matrix of dimension $M_2 \times M_1$, constructed from the normalized transmission matrix $t$, whose elements satisfy ${\rm Var}(t_{mn}) = 1/M_2$. We focus on the regime $M_1 \gg M_2$, with $M_1 \in [100,1000]\cap\mathbb{Z}$ and $M_2 \in [1,5]\cap\mathbb{Z}$. Each sub-matrix $A$ is formed by randomly choosing $M_1$ input channels and $M_2$ output target positions $m$ in transmission.

For every sub-matrix size $M_2 \times M_1$, we compute the eigenvalues from
\begin{equation}
    A^\dagger A V_\alpha = \lambda_\alpha V_\alpha,
\end{equation}
and determine the probability density $P_{\rm max}(\lambda_{\max})$, the mean $\langle \lambda_{\max}\rangle$, and the standard deviation
\begin{equation}
    \Delta \lambda_{\max} = \sqrt{\langle \lambda_{\max}^2 \rangle - \langle \lambda_{\max}\rangle^2}
\end{equation}
of the largest eigenvalue. This procedure yields the statistical distribution of the enhancement factor for focusing light through a scattering slab. We apply it to transmission matrices \(t_{mn}\) of slabs with \(k_0 W = 6{,}000\), \(\Sigma = 1.45, 0.65, 0.60\) (the normalized transport mean free path of $k_0l_t = 3, 18, 20.6$), \(n_0 = n_1 = n_2 = 1.5\), and normalized thicknesses \(k_0 L \in \{1, 50, 200\}\).

\section{Supplementary Information}

The following sections provide supplementary information for “Largest eigenvalue statistics of wavefront shaping in complex scattering media.” The first subsection provides additional details of the analytical approach based on random matrix theory. Subsections 2–4 present additional numerical and experimental results supporting the main text.

\subsection{Analytical approach via random matrix theory}

In this subsection, we provide some details about the analytical results provided in the main text. Our starting point is the joint probability distribution function (JPDF) of the $M_2$ real and nonzero eigenvalues given for the focusing operator $A^\dagger A$ in Eq.~\ref{Qmax}, which we recall here
\begin{equation} \label{JPDF_SM}
P(\lambda_1,\dots,\lambda_{M_2}) = \frac{1}{Z} \prod_{i=1}^{M_2} \lambda_i^{\nu}\, e^{-M_2 \lambda_i} \prod_{i<j} (\lambda_i - \lambda_j)^2 \;,
\end{equation}
where $\nu = M_1 - M_2 >0$. The normalization factor can be explicitly computed, leading to
\begin{equation} \label{Z_SM}
Z = M_2^{-M_1 M_2} \prod_{k=0}^{M_2-1} [(k+1)!(k+\nu)!] \;.
\end{equation}

\subsubsection{Average density}

\begin{figure*}[t]
	\centering
	\includegraphics[width=\linewidth]{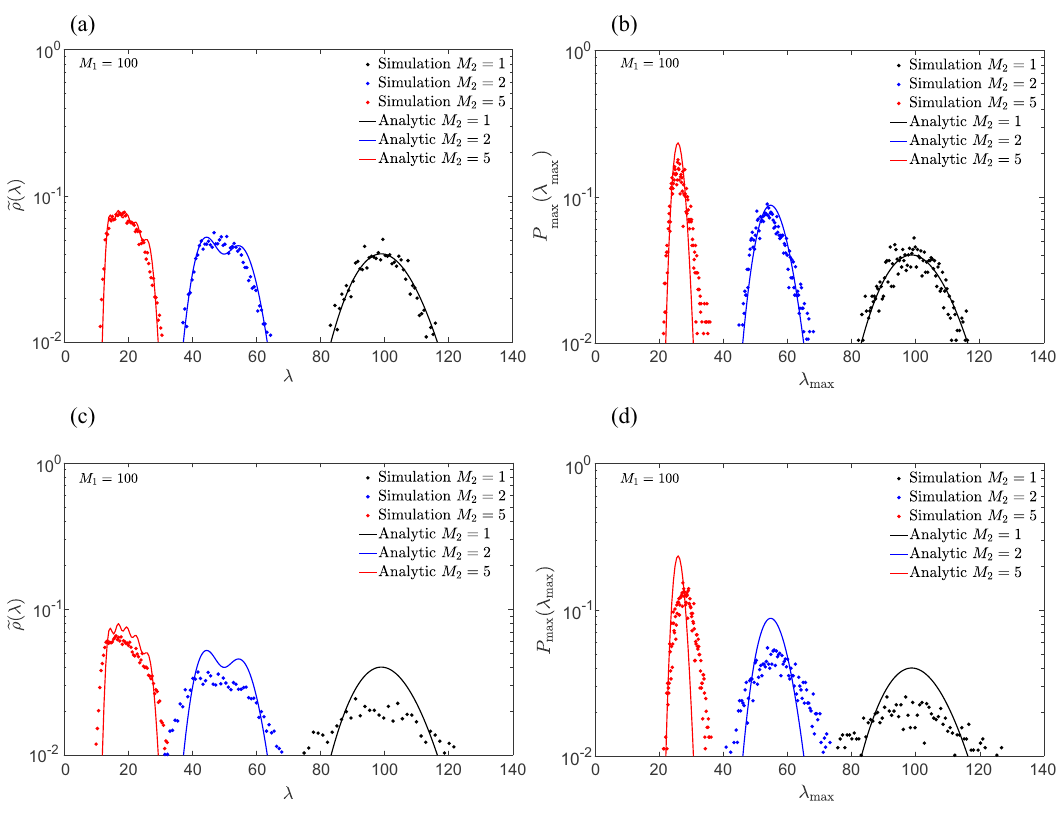}
	\caption{(a) Eigenvalue distributions $\widetilde{\rho}(\lambda)$ and (b) largest eigenvalue distributions $P_{\rm max}(\lambda_{\rm max})$ for $M_1 = 100$ and $M_2 = 1, 2, 5$ for a diffusive slab with normalized thickness $k_0L = 1$ and normalized transport mean free path $k_0l_t = 20.6$. For this case, the LW–random-matrix predictions for $\widetilde{\rho}(\lambda)$ [Eq.~(\ref{CD})] and for $P_{\rm max}(\lambda_{\rm max}) = Q'_{\max}(\lambda_{\max})$ [Eq.~(\ref{Qmax_expl1})] are in good agreement with the wave-propagation simulations. In contrast, for a thicker slab with $k_0L = 200$ and $k_0l_t = 20.6$, we find clear deviations between the LW–based predictions and the simulations for both $\widetilde{\rho}(\lambda)$ (c) and $P_{\rm max}(\lambda_{\rm max})$ (d).}
	\label{figureS2}
\end{figure*}

The density of nonzero eigenvalues $\tilde \rho(\lambda)$ is defined as
\begin{eqnarray} \label{tilde_rho_SM1}
\tilde \rho(\lambda) = \sum_{i=1}^{M_2} \langle \delta(\lambda-\lambda_i)\rangle
\end{eqnarray}
where $\delta(x)$ denotes the Dirac delta function and with the notation $\langle \cdots \rangle$ referring to an average over the random eigenvalues $\lambda_1, \cdots , \lambda_{M_2}$ distributed according to the JPDF in (\ref{JPDF_SM}). By inserting this explicit expression for the JPDF in Eq. (\ref{tilde_rho_SM1}) and using its invariance under permutation of the $\lambda_i$'s we see that computing $\tilde \rho(\lambda)$ amounts to perform the following multiple integral
\begin{eqnarray} \label{tilde_rho_SM2}
\tilde \rho(\lambda) = M_2\, \int_0^\infty d\lambda_2 \cdots \int_0^\infty d\lambda_{M_2} P(\lambda, \lambda_2, \cdots, \lambda_{M_2}) \;.
\end{eqnarray}
It turns out that this multiple integral can be performed explicitly using standard tools in random matrix theory (RMT), e.g., orthogonal polynomials, leading to the result given in Eq.~\ref{density}. In fact, in this formula~\ref{density}, the sum over $k$ can be explicitly computed using the Christoffel-Darboux relation 
\begin{eqnarray} \label{CD}
\tilde \rho(\lambda) = \frac{M_2!M_2^{\nu+1 }\lambda^\nu e^{-M_2 \lambda} }{(M_2+\nu-1)!} 
&&\Bigg[L_{M_2-1}^{(\nu)}(M_2 \lambda)L_{M_2-1}^{(\nu+1)}(M_2 \lambda) \nonumber \\
&&\hspace*{-0.5cm}-L_{M_2}^{(\nu)}(M_2 \lambda)L_{M_2-2}^{(\nu+1)}(M_2 \lambda) \Bigg] \;,
\end{eqnarray}
where $L_k^{(\nu)}(\lambda)$ is the generalized Laguerre polynomial of index $\nu$ and degree $k$, with the convention $L_{-1}^{\nu+1}(x) = 0$ for all $x$.
This density has a nontrivial structure exhibiting $M_2$ local maxima, corresponding to the typical location of the $M_2$ nonzero eigenvalues of the Anti-Wishart matrix $A^\dagger A$. Its asymptotic behaviors for small and large $\lambda$
 are given by
 \begin{eqnarray} \label{asympt_rhotilde}
 \tilde \rho(\lambda) \approx 
 \begin{cases}
 & C_0\, \lambda^\nu \quad, \hspace*{2.7cm} \lambda \to 0 \, \\
 & \\ 
 & C_\infty \, \lambda^{\nu + 2 M_2 -2}\, e^{-M_2\, \lambda} \quad, \quad \lambda \to \infty \;, 
 \end{cases}
 \end{eqnarray}
where the amplitudes $C_0$ and $C_\infty$ are given by
\begin{eqnarray} \label{CO_Cinf}
&&C_0 = \frac{M_2^{1+\nu}(M_2+\nu)!}{(M_2-1)!\nu ! (\nu+1)!} \;, \label{C0}\\
&&C_\infty = \frac{M_2^{\nu + 2 M_2-1}}{(M_2+\nu-1)!(M_2-1)!} \;.\label{Cinf}
\end{eqnarray}

A plot of the density given $\tilde \rho(\lambda)$ in Eq. (\ref{CD}) vs $\lambda$ is shown in Fig.~\ref{figureS1} (a) for $M_1 = 100$ and different values of $M_2 =1,2,5$ along with numerical results obtained by direct diagonalization of LW matrices, demonstrating a good agreement. For $M_2 = 2$ and $M_2 = 5$, we observe two and five local maxima, respectively. These peaks reflect the typical positions of the corresponding nonzero eigenvalues of $A^\dagger A$. Because of eigenvalue repulsion, these eigenvalues form a relatively rigid spectrum: their individual values fluctuate, but their locations remain sharply concentrated around well-defined mean positions. Consequently, the probability density develops distinct peaks near the characteristic positions of the first, second, third, and so on, eigenvalues. As $M_2$ becomes larger, these individual peaks gradually merge, and in the asymptotic limit of large $M_2$ (and $M_1$), the spectrum converges to the Mar\v{c}enko-Pastur distribution.

\begin{figure*}[t]
	\centering
	\includegraphics[width=\linewidth]{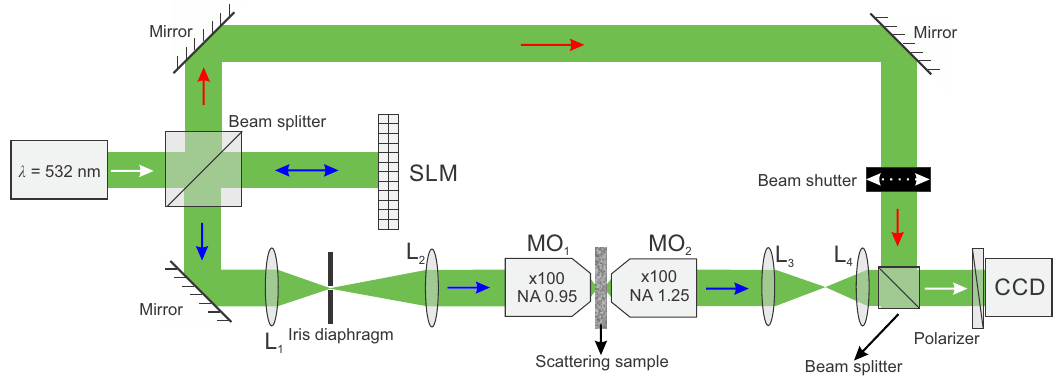}
	\caption{Schematic of the interferometric setup. A reflective phase-only spatial light modulator (SLM) shapes the phase of a monochromatic laser beam ($\lambda_0$ = 532 nm). Using the SLM and a CCD camera, the field transmission matrix of the scattering sample is measured in k-space. ${\rm MO}_{1,2}$: microscope objectives; ${\rm L}_{1-4}$: lenses. Reproduced from Ref.~\cite{yilmaz2021customizing} under the terms of the CC BY 4.0 license.}
	\label{figureS3}
\end{figure*}

\subsubsection{Cumulative distribution function of $\lambda_{\max}$} \label{sec:lmax}

As explained in the main text, the cumulative distribution function (CDF) of $Q_{\max}(x)={\rm Prob}(\lambda_{\max} \leq x) = {\rm Prob}{(\lambda_1\leq x, \cdots, \lambda_{M_2}\leq x)}$ can be written as the following multiple integral
\begin{eqnarray} \label{CDF_first}
Q_{\max}(x) = \int_0^x d \lambda_1 \cdots \int_0^x d \lambda_{M_2} P(\lambda_1,\dots,\lambda_{M_2}) \;.
\end{eqnarray}
This multiple integral can be performed explicitly using the standard Cauchy-Binet formula \cite{forrester2010log}, leading to 
\begin{equation} \label{Qmax_SM}
Q_{\max}(x) = \frac{M_2!}{Z} \, \det \mathbb{M}, \quad
\mathbb{M}_{i,j} = \int_0^x \lambda^{i+j-2+\nu} e^{-M_2 \lambda} \, d\lambda \;,
\end{equation}
with $i,j = 1, \cdots, M_2$. We recall that $Z$ is given in Eq. (\ref{Z_SM}). By expressing the integral over $\lambda$ that enters the definition of the matrix elements $\mathbb{M}_{i,j}$ in (\ref{Qmax_SM}) in terms of the incomplete gamma function $\gamma(s,z) = \int_0^z t^{s-1} e^{-t} \, dt$ and performing standard manipulations of the determinant yields the expression given in Eq.~\ref{Qmax}. In fact, since $\nu = M_1 - M_2 >0$ is an integer, this integral over $\lambda$ can be performed explicitly, leading to
\begin{eqnarray} \label{integral}
\mathbb{M}_{i,j} &=& \int_0^x \lambda^{i+j-2+\nu} e^{-M_2 \lambda} \, d\lambda \\
&=& \frac{(i+j-2+\nu)!}{M_2^{i+j-1+\nu}} \;\Bigg( 1 - e^{-M_2 x} \sum_{k=0}^{i+j-2+\nu} \frac{(M_2 x)^k}{k!} \Bigg),  \nonumber
\end{eqnarray}
which holds provided $i,j$ and $\nu = M_1 - M_2$ are integers (which is indeed the case here). Injecting this in Eq. (\ref{Qmax_SM}), one finds the fairly explicit expression
\begin{widetext}
\begin{equation} \label{Qmax_expl1}
Q_{\max}(x) = \frac{M_2!}{\prod_{k=0}^{M_2-1} (k+1)! \, (k+\nu)!}\, 
\det_{1\leq i,j \leq M_2} \Bigg[(i+j-2+\nu)! \;\Bigg( 1 - e^{-M_2 x} \sum_{k=0}^{i+j-2+\nu} \frac{(M_2 x)^k}{k!} \Bigg) \;,
\Bigg] \quad, \; \nu = M_1 - M_2 \;.
\end{equation}
\end{widetext}
Note that for $M_2 = 1$ one finds the simple result $P_{\max}(x) = d Q_{\max}(x)/dx = x^{M_1-1}\, e^{-x}/(M_1-1)!$, as in \cite{PaniaguaDiaz2023}, but already for $M_2=2$ the expression is much more complicated.   

From this explicit expression, one can obtain the asymptotic behaviors of $Q_{\max(x)}$. As $x \to 0$ it behaves as $Q_{\max}(x) \propto x^{M_1 M_2}$ while $1-Q_{\max}(x) \to q_{M_1, M_2}(x)e^{-M_2 x}$ as $x \to \infty$, where $q_{M_1,M_2}(x)$ is a polynomial of order $M_1 +M_2-2$. 

A plot of the density given $P_{\max}(\lambda_{\max}) = Q'_{\max}(\lambda_{\max})$ from Eq. (\ref{Qmax_expl1}) vs $\lambda_{\max}$ is shown in Fig.~\ref{figureS1} (b) for $M_1 = 100$ and different values of $M_2 =1,2,5$ along with numerical results obtained by direct diagonalization of LW matrices, demonstrating a good agreement. 

\subsubsection{The limit of large $M_1 \gg M_2$}

In the limit where $M_1 \gg M_2$ one can show that the complex Wishart ensemble described by (\ref{JPDF_SM}) converges, after centering around $M_1/M_2$ and rescaling by $M_2/\sqrt{2 M_1}$, to the Gaussian Unitary Ensemble (GUE) that describes Hermitian matrices with independent (complex) Gaussian entries. Consequently for large $M_1 \gg  M_ 2$ one finds
\begin{eqnarray} \label{large_nu}
\langle \lambda_{\max} \rangle \approx \frac{1}{M_2}\left( M_1 +\sqrt{2 \,M_1} \, \langle \mu_{\max}\rangle \right) \;,
\end{eqnarray}
where $\mu_{\max}$ denotes the largest eigenvalue of a GUE matrix of size $M_2 \times M_2$. Its CDF ${\cal Q}_{\max}(x) = {\rm Prob.}(\mu_{\max} \leq x)$ can be computed as explained above, but transposed to the case of GUE. It reads
\begin{equation} \label{CDF_GUE}
{\cal Q}_{\max}(x) = \frac{1}{\tilde Z} {\det} \, \widetilde{\mathbb{M}} \;, \; \widetilde{\mathbb{M}}_{i,j} = {\int_{-\infty}^x \mu^{i+j-2}\,e^{-\mu^2}d\mu}\;,
\end{equation}
with $\tilde Z = \det^{M_2}_{i,j=1}  \int_{-\infty}^\infty \mu^{i+j-2}\,e^{-\mu^2}d\mu$. From this expression for the CDF in (\ref{CDF_GUE}) one can obtain the average $\langle \mu_{\max}\rangle = \int_{-\infty}^\infty x {\cal Q}'_{\max}(x) dx$ yielding in particular 
\begin{eqnarray} \label{mu_max}
\langle \mu_{\max} \rangle =
\begin{cases}
& 0 \quad , \quad \hspace*{0.75cm} M_2 = 1 \;,\\
& \\
& \sqrt{\dfrac{2}{\pi}} \quad, \quad \hspace*{0.3cm} M_2 = 2 \;,\\
& \\
& \dfrac{27}{8 \sqrt{2 \pi}} \quad, \quad M_2 = 3 \;,
\end{cases} 
\end{eqnarray}
as given in the text. 

In this limit $M_1 \gg M_2$ one can also compute the behavior of the standard deviation. Indeed from the convergence of the complex LW ensemble to the GUE (of size $M_2 \times M_2$), one has the identity in law
\begin{eqnarray} \label{id_law}
\lambda_{\max} \approx \frac{1}{M_2}(M_1 + \sqrt{2 M_1} \mu_{\max}) \;,
\end{eqnarray}
where, as before, $\mu_{\max}$, which depends only $M_2$ and not on $M_1$, denotes the largest eigenvalue of a GUE matrix of size $M_2 \times M_2$. This identity (\ref{id_law}) allows to obtain straightforwardly the large $M_1$ behavior of the standard deviation $\Delta \lambda_{\max}$ as given in the text, namely 
\begin{equation} \label{std_dev_sm}
\Delta \lambda_{\max} \approx {\frac{\sqrt{2 M_1}}{M_2}} \Delta \mu_{\max} \;,
\end{equation}
where $\Delta \mu_{\max}$, which depends only on $M_2$, is the standard deviation of the largest eigenvalue of a GUE matrix of size $M_2 \times M_2$. It can be computed from the CDF of $\mu_{\max}$ given in Eq. (\ref{CDF_GUE}). In particular, one gets
\begin{eqnarray} \label{std_GUE}
\left(\Delta \mu_{\max}\right)^2 =
\begin{cases}
& \dfrac{1}{2} \quad, \quad \hspace*{2.6cm} M_2 = 1 \;, \\
& \\
& 1 - \dfrac{2}{\pi} \quad, \quad \hspace*{2cm} M_2 = 2\;, \\
& \\
& \dfrac{3}{2} + \dfrac{9}{128\pi}(16 \sqrt{3}-81) \;, \; M_2=3 \;.
\end{cases} \nonumber \\
\end{eqnarray}
These asymptotic behaviors in Eqs. (\ref{large_nu})-(\ref{mu_max}) and (\ref{std_dev_sm})-(\ref{std_GUE})
are compared to numerical and exact results in Fig.~\ref{figure3}, showing an excellent agreement between the different estimations. 

\begin{figure*}[t]
	\centering
	\includegraphics[width=\linewidth]{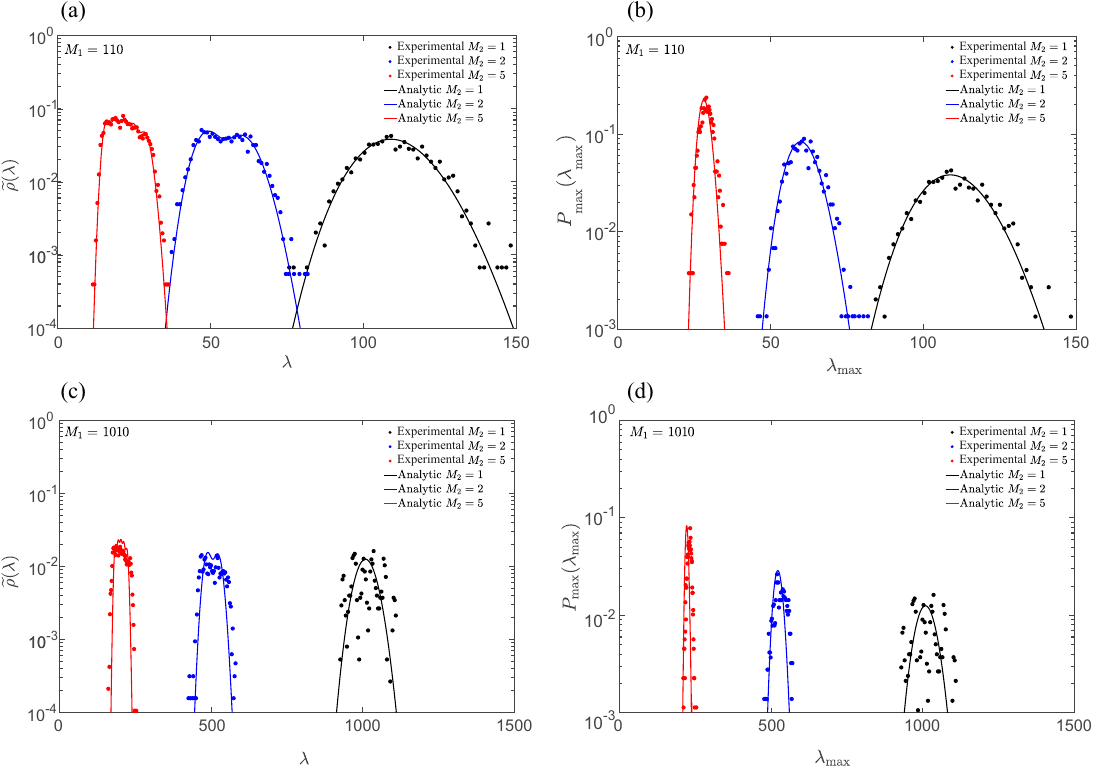}
	\caption{{(a) Eigenvalue distributions $\widetilde{\rho}(\lambda)$ and (b) largest eigenvalue distributions $P_{\rm max}(\lambda_{\rm max})$ for $M_1 = 110$ and $M_2 = 1, 2, 5$ for the experimental diffusive slab. For this case, the LW–random-matrix predictions for $\widetilde{\rho}(\lambda)$ [Eq.~(\ref{CD})] and for $P_{\rm max}(\lambda_{\rm max}) = Q'_{\max}(\lambda_{\max})$ [Eq.~(\ref{Qmax_expl1})] are in good agreement with the experimental results. In contrast, for a larger $M_1 = 1010$, we find clear deviations between the LW–based predictions and the experiments for both $\widetilde{\rho}(\lambda)$ (c) and $P_{\rm max}(\lambda_{\rm max})$ (d).}}
	\label{figureS4}
\end{figure*}

\subsection{Eigenvalue probability densities from wave-propagation simulations}

Fig.~\ref{figureS2} shows the eigenvalue distributions $\widetilde{\rho}(\lambda)$ and the largest eigenvalue distributions $P_{\rm max}(\lambda_{\rm max})$ obtained from LW–random-matrix theory and from wave-propagation simulations for diffusive slabs with $M_1 = 100$ and $M_2 = 1,2,5$. For the thin slab with $k_0L = 1$ and $k_0l_t = 20.6$ [panels (a,b)], the LW predictions accurately reproduce both the bulk eigenvalue density and the statistics of $\lambda_{\rm max}$, indicating that long-range mesoscopic correlations are negligible. In contrast, for the thicker scattering slab with $k_0L = 200$ and $k_0l_t = 20.6$ [panels (c,d)], clear, systematic deviations emerge in both $\widetilde{\rho}(\lambda)$ and $P_{\rm max}(\lambda_{\rm max})$. These deviations signal the breakdown of the uncorrelated-random-matrix assumption and reflect the growing impact of long-range mesoscopic correlations, which reshape both the eigenvalue distribution and the distribution of the largest eigenvalues.

\subsection{Experimental setup and eigenvalue probability densities}

The experimental setup used to measure the transmission matrix of the diffusive sample is shown in Fig.~\ref{figureS3}. Fig.~\ref{figureS4} presents the eigenvalue densities $\widetilde{\rho}(\lambda)$ and largest eigenvalue distributions $P_{\rm max}(\lambda_{\rm max})$ obtained from LW–random-matrix theory and from experiment. For $M_1 = 110$ [panels (a,b)], the LW predictions faithfully capture both the bulk eigenvalue density and the statistics of $\lambda_{\rm max}$, indicating that long-range mesoscopic correlations are negligible. In contrast, for $M_1 = 1010$ [panels (c,d)], pronounced and systematic deviations appear in both $\widetilde{\rho}(\lambda)$ and $P_{\rm max}(\lambda_{\rm max})$. These deviations mark the breakdown of the uncorrelated random-matrix description and reveal the increasing influence of long-range mesoscopic correlations, which alter both the overall eigenvalue density and the distribution of the largest eigenvalues.

\begin{figure}[t]
	\centering
	\includegraphics[width=\linewidth]{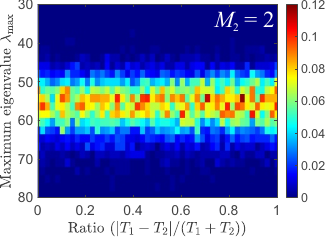}
	\caption{Experimental joint probability density (JPD) for two-target focusing ($M_{1}=100$, $M_{2}=2$): there is no discernible dependence between $\lambda_{\max}$ and the output intensity ratio, so high-$\lambda_{\max}$ events also occur when the two target intensities are comparable (near-equal split).}
	\label{figureS5}
\end{figure}

\subsection{Enhancement factor and intensity balance in two-target focusing}

In multipoint focusing experiments, the goal is often not only to maximize the enhancement factor but also to obtain similar intensities at the target positions. It is therefore important to check whether the intensity balance between targets is correlated with the achieved enhancement. To this end, we evaluated the experimental joint probability density of the largest eigenvalue \(\lambda_{\rm max}\) and the intensity imbalance for two-target focusing \((M_2 = 2)\), defined as $R \equiv \frac{\lvert T_1 - T_2 \rvert}{T_1 + T_2}$, where \(T_1\) and \(T_2\) are the intensities at the two target points (Fig.~\ref{figureS5}). The analysis shows no discernible correlation between \(\lambda_{\rm max}\) and \(R\). In other words, even for large enhancement factors, one can observe either perfectly balanced intensities (\(R = 0\), \(T_1 = T_2\)) or completely unbalanced ones (\(R = 1\), \(T_1 = 0\) or \(T_2 = 0\)) with comparable likelihood.

\section{Data availability}
The raw transmission-matrix datasets generated in this study are not deposited in a public repository because of their large file size. These data are available from the corresponding authors upon request. Requests should be directed to either corresponding author and will normally be addressed within eight weeks. There are no restrictions on who may request the data or on their use for research purposes. Once provided, the data will remain available to the requester without a time limit.

\section{Acknowledgements}
We thank Hui Cao for supporting the original transmission-matrix measurements and for providing access to the dataset, as well as for carefully reading the manuscript, Arthur Goetschy for insightful discussions, and Ilgın Yağmur Şen for initial contributions. H. Y. thanks Soner Albayrak for helpful discussions.

\section{Funding}
G. S. acknowledges support from ANR Grant No. ANR-23-CE30-0020-01 EDIPS.

\section{Author contributions}
H.Y. conceived the project, performed the experiments and numerical simulations, and analyzed the data. G.S. conceived the analytical approach and carried out the analytical calculations. Both authors interpreted the results, wrote the manuscript, and approved the final version.

\section{Competing interests}
The authors declare no competing interests.

% Bibliography
\bibliography{focusing_statistics_references}

\begin{thebibliography}{97}%
\makeatletter
\providecommand \@ifxundefined [1]{%
 \@ifx{#1\undefined}
}%
\providecommand \@ifnum [1]{%
 \ifnum #1\expandafter \@firstoftwo
 \else \expandafter \@secondoftwo
 \fi
}%
\providecommand \@ifx [1]{%
 \ifx #1\expandafter \@firstoftwo
 \else \expandafter \@secondoftwo
 \fi
}%
\providecommand \natexlab [1]{#1}%
\providecommand \enquote  [1]{``#1''}%
\providecommand \bibnamefont  [1]{#1}%
\providecommand \bibfnamefont [1]{#1}%
\providecommand \citenamefont [1]{#1}%
\providecommand \href@noop [0]{\@secondoftwo}%
\providecommand \href [0]{\begingroup \@sanitize@url \@href}%
\providecommand \@href[1]{\@@startlink{#1}\@@href}%
\providecommand \@@href[1]{\endgroup#1\@@endlink}%
\providecommand \@sanitize@url [0]{\catcode `\\12\catcode `\$12\catcode `\&12\catcode `\#12\catcode `\^12\catcode `\_12\catcode `\%12\relax}%
\providecommand \@@startlink[1]{}%
\providecommand \@@endlink[0]{}%
\providecommand \url  [0]{\begingroup\@sanitize@url \@url }%
\providecommand \@url [1]{\endgroup\@href {#1}{\urlprefix }}%
\providecommand \urlprefix  [0]{URL }%
\providecommand \Eprint [0]{\href }%
\providecommand \doibase [0]{https://doi.org/}%
\providecommand \selectlanguage [0]{\@gobble}%
\providecommand \bibinfo  [0]{\@secondoftwo}%
\providecommand \bibfield  [0]{\@secondoftwo}%
\providecommand \translation [1]{[#1]}%
\providecommand \BibitemOpen [0]{}%
\providecommand \bibitemStop [0]{}%
\providecommand \bibitemNoStop [0]{.\EOS\space}%
\providecommand \EOS [0]{\spacefactor3000\relax}%
\providecommand \BibitemShut  [1]{\csname bibitem#1\endcsname}%
\let\auto@bib@innerbib\@empty
%</preamble>
\bibitem [{\citenamefont {Vellekoop}\ and\ \citenamefont {Mosk}(2007)}]{2007_Vellekoop}%
  \BibitemOpen
  \bibfield  {author} {\bibinfo {author} {\bibfnamefont {I.~M.}\ \bibnamefont {Vellekoop}}\ and\ \bibinfo {author} {\bibfnamefont {A.~P.}\ \bibnamefont {Mosk}},\ }\bibfield  {title} {\bibinfo {title} {Focusing coherent light through opaque strongly scattering media},\ }\href@noop {} {\bibfield  {journal} {\bibinfo  {journal} {Opt. Lett.}\ }\textbf {\bibinfo {volume} {32}},\ \bibinfo {pages} {2309} (\bibinfo {year} {2007})}\BibitemShut {NoStop}%
\bibitem [{\citenamefont {Mosk}\ \emph {et~al.}(2012)\citenamefont {Mosk}, \citenamefont {Lagendijk}, \citenamefont {Lerosey},\ and\ \citenamefont {Fink}}]{2012_Mosk_NatPhoton_R}%
  \BibitemOpen
  \bibfield  {author} {\bibinfo {author} {\bibfnamefont {A.~P.}\ \bibnamefont {Mosk}}, \bibinfo {author} {\bibfnamefont {A.}~\bibnamefont {Lagendijk}}, \bibinfo {author} {\bibfnamefont {G.}~\bibnamefont {Lerosey}},\ and\ \bibinfo {author} {\bibfnamefont {M.}~\bibnamefont {Fink}},\ }\bibfield  {title} {\bibinfo {title} {Controlling waves in space and time for imaging and focusing in complex media},\ }\href@noop {} {\bibfield  {journal} {\bibinfo  {journal} {Nat. Photonics}\ }\textbf {\bibinfo {volume} {6}},\ \bibinfo {pages} {283} (\bibinfo {year} {2012})}\BibitemShut {NoStop}%
\bibitem [{\citenamefont {Yu}\ \emph {et~al.}(2015)\citenamefont {Yu}, \citenamefont {Park}, \citenamefont {Lee}, \citenamefont {Yoon}, \citenamefont {Kim}, \citenamefont {Lee},\ and\ \citenamefont {Park}}]{2015_Park_R}%
  \BibitemOpen
  \bibfield  {author} {\bibinfo {author} {\bibfnamefont {H.}~\bibnamefont {Yu}}, \bibinfo {author} {\bibfnamefont {J.}~\bibnamefont {Park}}, \bibinfo {author} {\bibfnamefont {K.}~\bibnamefont {Lee}}, \bibinfo {author} {\bibfnamefont {J.}~\bibnamefont {Yoon}}, \bibinfo {author} {\bibfnamefont {K.}~\bibnamefont {Kim}}, \bibinfo {author} {\bibfnamefont {S.}~\bibnamefont {Lee}},\ and\ \bibinfo {author} {\bibfnamefont {Y.}~\bibnamefont {Park}},\ }\bibfield  {title} {\bibinfo {title} {Recent advances in wavefront shaping techniques for biomedical applications},\ }\href@noop {} {\bibfield  {journal} {\bibinfo  {journal} {Curr. Appl. Phys.}\ }\textbf {\bibinfo {volume} {15}},\ \bibinfo {pages} {632} (\bibinfo {year} {2015})}\BibitemShut {NoStop}%
\bibitem [{\citenamefont {Vellekoop}(2015)}]{2015_Vellekoop_OptExpress}%
  \BibitemOpen
  \bibfield  {author} {\bibinfo {author} {\bibfnamefont {I.~M.}\ \bibnamefont {Vellekoop}},\ }\bibfield  {title} {\bibinfo {title} {Feedback-based wavefront shaping},\ }\href@noop {} {\bibfield  {journal} {\bibinfo  {journal} {Opt. Express}\ }\textbf {\bibinfo {volume} {23}},\ \bibinfo {pages} {12189} (\bibinfo {year} {2015})}\BibitemShut {NoStop}%
\bibitem [{\citenamefont {Horstmeyer}\ \emph {et~al.}(2015)\citenamefont {Horstmeyer}, \citenamefont {Ruan},\ and\ \citenamefont {Yang}}]{2015_Yang_NatPhoton}%
  \BibitemOpen
  \bibfield  {author} {\bibinfo {author} {\bibfnamefont {R.}~\bibnamefont {Horstmeyer}}, \bibinfo {author} {\bibfnamefont {H.}~\bibnamefont {Ruan}},\ and\ \bibinfo {author} {\bibfnamefont {C.}~\bibnamefont {Yang}},\ }\bibfield  {title} {\bibinfo {title} {Guidestar-assisted wavefront-shaping methods for focusing light into biological tissue},\ }\href@noop {} {\bibfield  {journal} {\bibinfo  {journal} {Nat. Photonics}\ }\textbf {\bibinfo {volume} {9}},\ \bibinfo {pages} {563} (\bibinfo {year} {2015})}\BibitemShut {NoStop}%
\bibitem [{\citenamefont {Rotter}\ and\ \citenamefont {Gigan}(2017)}]{2017_Rotter_RMP_R}%
  \BibitemOpen
  \bibfield  {author} {\bibinfo {author} {\bibfnamefont {S.}~\bibnamefont {Rotter}}\ and\ \bibinfo {author} {\bibfnamefont {S.}~\bibnamefont {Gigan}},\ }\bibfield  {title} {\bibinfo {title} {Light fields in complex media: mesoscopic scattering meets wave control},\ }\href@noop {} {\bibfield  {journal} {\bibinfo  {journal} {Rev. Mod. Phys.}\ }\textbf {\bibinfo {volume} {89}},\ \bibinfo {pages} {015005} (\bibinfo {year} {2017})}\BibitemShut {NoStop}%
\bibitem [{\citenamefont {Gigan}\ \emph {et~al.}(2022)\citenamefont {Gigan}, \citenamefont {Katz}, \citenamefont {De~Aguiar}, \citenamefont {Andresen}, \citenamefont {Aubry}, \citenamefont {Bertolotti}, \citenamefont {Bossy}, \citenamefont {Bouchet}, \citenamefont {Brake}, \citenamefont {Brasselet} \emph {et~al.}}]{gigan2022roadmap}%
  \BibitemOpen
  \bibfield  {author} {\bibinfo {author} {\bibfnamefont {S.}~\bibnamefont {Gigan}}, \bibinfo {author} {\bibfnamefont {O.}~\bibnamefont {Katz}}, \bibinfo {author} {\bibfnamefont {H.~B.}\ \bibnamefont {De~Aguiar}}, \bibinfo {author} {\bibfnamefont {E.~R.}\ \bibnamefont {Andresen}}, \bibinfo {author} {\bibfnamefont {A.}~\bibnamefont {Aubry}}, \bibinfo {author} {\bibfnamefont {J.}~\bibnamefont {Bertolotti}}, \bibinfo {author} {\bibfnamefont {E.}~\bibnamefont {Bossy}}, \bibinfo {author} {\bibfnamefont {D.}~\bibnamefont {Bouchet}}, \bibinfo {author} {\bibfnamefont {J.}~\bibnamefont {Brake}}, \bibinfo {author} {\bibfnamefont {S.}~\bibnamefont {Brasselet}}, \emph {et~al.},\ }\bibfield  {title} {\bibinfo {title} {Roadmap on wavefront shaping and deep imaging in complex media},\ }\href@noop {} {\bibfield  {journal} {\bibinfo  {journal} {J. Phys. Photonics}\ }\textbf {\bibinfo {volume} {4}},\ \bibinfo {pages} {042501} (\bibinfo {year} {2022})}\BibitemShut {NoStop}%
\bibitem [{\citenamefont {Cao}\ \emph {et~al.}(2022)\citenamefont {Cao}, \citenamefont {Mosk},\ and\ \citenamefont {Rotter}}]{cao2022shaping}%
  \BibitemOpen
  \bibfield  {author} {\bibinfo {author} {\bibfnamefont {H.}~\bibnamefont {Cao}}, \bibinfo {author} {\bibfnamefont {A.~P.}\ \bibnamefont {Mosk}},\ and\ \bibinfo {author} {\bibfnamefont {S.}~\bibnamefont {Rotter}},\ }\bibfield  {title} {\bibinfo {title} {Shaping the propagation of light in complex media},\ }\href@noop {} {\bibfield  {journal} {\bibinfo  {journal} {Nat. Phys.}\ }\textbf {\bibinfo {volume} {18}},\ \bibinfo {pages} {994} (\bibinfo {year} {2022})}\BibitemShut {NoStop}%
\bibitem [{\citenamefont {Lerosey}\ \emph {et~al.}(2007)\citenamefont {Lerosey}, \citenamefont {De~Rosny}, \citenamefont {Tourin},\ and\ \citenamefont {Fink}}]{lerosey2007focusing}%
  \BibitemOpen
  \bibfield  {author} {\bibinfo {author} {\bibfnamefont {G.}~\bibnamefont {Lerosey}}, \bibinfo {author} {\bibfnamefont {J.}~\bibnamefont {De~Rosny}}, \bibinfo {author} {\bibfnamefont {A.}~\bibnamefont {Tourin}},\ and\ \bibinfo {author} {\bibfnamefont {M.}~\bibnamefont {Fink}},\ }\bibfield  {title} {\bibinfo {title} {Focusing beyond the diffraction limit with far-field time reversal},\ }\href@noop {} {\bibfield  {journal} {\bibinfo  {journal} {Science}\ }\textbf {\bibinfo {volume} {315}},\ \bibinfo {pages} {1120} (\bibinfo {year} {2007})}\BibitemShut {NoStop}%
\bibitem [{\citenamefont {Vellekoop}\ and\ \citenamefont {Mosk}(2008{\natexlab{a}})}]{2008_Vellekoop_PRL}%
  \BibitemOpen
  \bibfield  {author} {\bibinfo {author} {\bibfnamefont {I.~M.}\ \bibnamefont {Vellekoop}}\ and\ \bibinfo {author} {\bibfnamefont {A.~P.}\ \bibnamefont {Mosk}},\ }\bibfield  {title} {\bibinfo {title} {Universal optimal transmission of light through disordered materials},\ }\href@noop {} {\bibfield  {journal} {\bibinfo  {journal} {Phys. Rev. Lett.}\ }\textbf {\bibinfo {volume} {101}},\ \bibinfo {pages} {120601} (\bibinfo {year} {2008}{\natexlab{a}})}\BibitemShut {NoStop}%
\bibitem [{\citenamefont {Vellekoop}\ \emph {et~al.}(2010)\citenamefont {Vellekoop}, \citenamefont {Lagendijk},\ and\ \citenamefont {Mosk}}]{2010_vellekoop_Nat._Photonics}%
  \BibitemOpen
  \bibfield  {author} {\bibinfo {author} {\bibfnamefont {I.~M.}\ \bibnamefont {Vellekoop}}, \bibinfo {author} {\bibfnamefont {A.}~\bibnamefont {Lagendijk}},\ and\ \bibinfo {author} {\bibfnamefont {A.}~\bibnamefont {Mosk}},\ }\bibfield  {title} {\bibinfo {title} {Exploiting disorder for perfect focusing},\ }\href@noop {} {\bibfield  {journal} {\bibinfo  {journal} {Nat. Photonics}\ }\textbf {\bibinfo {volume} {4}},\ \bibinfo {pages} {320} (\bibinfo {year} {2010})}\BibitemShut {NoStop}%
\bibitem [{\citenamefont {Davy}\ \emph {et~al.}(2012)\citenamefont {Davy}, \citenamefont {Shi},\ and\ \citenamefont {Genack}}]{davy2012focusing}%
  \BibitemOpen
  \bibfield  {author} {\bibinfo {author} {\bibfnamefont {M.}~\bibnamefont {Davy}}, \bibinfo {author} {\bibfnamefont {Z.}~\bibnamefont {Shi}},\ and\ \bibinfo {author} {\bibfnamefont {A.~Z.}\ \bibnamefont {Genack}},\ }\bibfield  {title} {\bibinfo {title} {Focusing through random media: Eigenchannel participation number and intensity correlation},\ }\href@noop {} {\bibfield  {journal} {\bibinfo  {journal} {Phys. Rev. B}\ }\textbf {\bibinfo {volume} {85}},\ \bibinfo {pages} {035105} (\bibinfo {year} {2012})}\BibitemShut {NoStop}%
\bibitem [{\citenamefont {Kim}\ \emph {et~al.}(2012)\citenamefont {Kim}, \citenamefont {Choi}, \citenamefont {Yoon}, \citenamefont {Choi}, \citenamefont {Kim}, \citenamefont {Park},\ and\ \citenamefont {Choi}}]{2012_Choi_Nat._Photonics}%
  \BibitemOpen
  \bibfield  {author} {\bibinfo {author} {\bibfnamefont {M.}~\bibnamefont {Kim}}, \bibinfo {author} {\bibfnamefont {Y.}~\bibnamefont {Choi}}, \bibinfo {author} {\bibfnamefont {C.}~\bibnamefont {Yoon}}, \bibinfo {author} {\bibfnamefont {W.}~\bibnamefont {Choi}}, \bibinfo {author} {\bibfnamefont {J.}~\bibnamefont {Kim}}, \bibinfo {author} {\bibfnamefont {Q.-H.}\ \bibnamefont {Park}},\ and\ \bibinfo {author} {\bibfnamefont {W.}~\bibnamefont {Choi}},\ }\bibfield  {title} {\bibinfo {title} {Maximal energy transport through disordered media with the implementation of transmission eigenchannels},\ }\href@noop {} {\bibfield  {journal} {\bibinfo  {journal} {Nat. Photonics}\ }\textbf {\bibinfo {volume} {6}},\ \bibinfo {pages} {581} (\bibinfo {year} {2012})}\BibitemShut {NoStop}%
\bibitem [{\citenamefont {Davy}\ \emph {et~al.}(2013)\citenamefont {Davy}, \citenamefont {Shi}, \citenamefont {Wang},\ and\ \citenamefont {Genack}}]{davy2013transmission}%
  \BibitemOpen
  \bibfield  {author} {\bibinfo {author} {\bibfnamefont {M.}~\bibnamefont {Davy}}, \bibinfo {author} {\bibfnamefont {Z.}~\bibnamefont {Shi}}, \bibinfo {author} {\bibfnamefont {J.}~\bibnamefont {Wang}},\ and\ \bibinfo {author} {\bibfnamefont {A.~Z.}\ \bibnamefont {Genack}},\ }\bibfield  {title} {\bibinfo {title} {Transmission statistics and focusing in single disordered samples},\ }\href@noop {} {\bibfield  {journal} {\bibinfo  {journal} {Opt. Express}\ }\textbf {\bibinfo {volume} {21}},\ \bibinfo {pages} {10367} (\bibinfo {year} {2013})}\BibitemShut {NoStop}%
\bibitem [{\citenamefont {Popoff}\ \emph {et~al.}(2014)\citenamefont {Popoff}, \citenamefont {Goetschy}, \citenamefont {Liew}, \citenamefont {Stone},\ and\ \citenamefont {Cao}}]{2014_Popoff_PRL}%
  \BibitemOpen
  \bibfield  {author} {\bibinfo {author} {\bibfnamefont {S.~M.}\ \bibnamefont {Popoff}}, \bibinfo {author} {\bibfnamefont {A.}~\bibnamefont {Goetschy}}, \bibinfo {author} {\bibfnamefont {S.~F.}\ \bibnamefont {Liew}}, \bibinfo {author} {\bibfnamefont {A.~D.}\ \bibnamefont {Stone}},\ and\ \bibinfo {author} {\bibfnamefont {H.}~\bibnamefont {Cao}},\ }\bibfield  {title} {\bibinfo {title} {Coherent control of total transmission of light through disordered media},\ }\href@noop {} {\bibfield  {journal} {\bibinfo  {journal} {Phys. Rev. Lett.}\ }\textbf {\bibinfo {volume} {112}},\ \bibinfo {pages} {133903} (\bibinfo {year} {2014})}\BibitemShut {NoStop}%
\bibitem [{\citenamefont {Ojambati}\ \emph {et~al.}(2016{\natexlab{a}})\citenamefont {Ojambati}, \citenamefont {Y{\i}lmaz}, \citenamefont {Lagendijk}, \citenamefont {Mosk},\ and\ \citenamefont {Vos}}]{ojambati2016coupling}%
  \BibitemOpen
  \bibfield  {author} {\bibinfo {author} {\bibfnamefont {O.~S.}\ \bibnamefont {Ojambati}}, \bibinfo {author} {\bibfnamefont {H.}~\bibnamefont {Y{\i}lmaz}}, \bibinfo {author} {\bibfnamefont {A.}~\bibnamefont {Lagendijk}}, \bibinfo {author} {\bibfnamefont {A.~P.}\ \bibnamefont {Mosk}},\ and\ \bibinfo {author} {\bibfnamefont {W.~L.}\ \bibnamefont {Vos}},\ }\bibfield  {title} {\bibinfo {title} {Coupling of energy into the fundamental diffusion mode of a complex nanophotonic medium},\ }\href@noop {} {\bibfield  {journal} {\bibinfo  {journal} {New J. Phys.}\ }\textbf {\bibinfo {volume} {18}},\ \bibinfo {pages} {043032} (\bibinfo {year} {2016}{\natexlab{a}})}\BibitemShut {NoStop}%
\bibitem [{\citenamefont {Hsu}\ \emph {et~al.}(2017)\citenamefont {Hsu}, \citenamefont {Liew}, \citenamefont {Goetschy}, \citenamefont {Cao},\ and\ \citenamefont {Stone}}]{2017_Wade_Nat._Phys.}%
  \BibitemOpen
  \bibfield  {author} {\bibinfo {author} {\bibfnamefont {C.~W.}\ \bibnamefont {Hsu}}, \bibinfo {author} {\bibfnamefont {S.~F.}\ \bibnamefont {Liew}}, \bibinfo {author} {\bibfnamefont {A.}~\bibnamefont {Goetschy}}, \bibinfo {author} {\bibfnamefont {H.}~\bibnamefont {Cao}},\ and\ \bibinfo {author} {\bibfnamefont {A.~D.}\ \bibnamefont {Stone}},\ }\bibfield  {title} {\bibinfo {title} {Correlation-enhanced control of wave focusing in disordered media},\ }\href@noop {} {\bibfield  {journal} {\bibinfo  {journal} {Nat. Phys.}\ }\textbf {\bibinfo {volume} {13}},\ \bibinfo {pages} {497} (\bibinfo {year} {2017})}\BibitemShut {NoStop}%
\bibitem [{\citenamefont {Yılmaz}\ \emph {et~al.}(2019)\citenamefont {Yılmaz}, \citenamefont {Hsu}, \citenamefont {Yamilov},\ and\ \citenamefont {Cao}}]{2019_Yilmaz_Nat._Photonics}%
  \BibitemOpen
  \bibfield  {author} {\bibinfo {author} {\bibfnamefont {H.}~\bibnamefont {Yılmaz}}, \bibinfo {author} {\bibfnamefont {C.~W.}\ \bibnamefont {Hsu}}, \bibinfo {author} {\bibfnamefont {A.}~\bibnamefont {Yamilov}},\ and\ \bibinfo {author} {\bibfnamefont {H.}~\bibnamefont {Cao}},\ }\bibfield  {title} {\bibinfo {title} {Transverse localization of transmission eigenchannels},\ }\href@noop {} {\bibfield  {journal} {\bibinfo  {journal} {Nat. Photonics}\ }\textbf {\bibinfo {volume} {13}},\ \bibinfo {pages} {352} (\bibinfo {year} {2019})}\BibitemShut {NoStop}%
\bibitem [{\citenamefont {Y{\i}lmaz}\ \emph {et~al.}(2019)\citenamefont {Y{\i}lmaz}, \citenamefont {Hsu}, \citenamefont {Goetschy}, \citenamefont {Bittner}, \citenamefont {Rotter}, \citenamefont {Yamilov},\ and\ \citenamefont {Cao}}]{2019_Yilmaz2019_PRL}%
  \BibitemOpen
  \bibfield  {author} {\bibinfo {author} {\bibfnamefont {H.}~\bibnamefont {Y{\i}lmaz}}, \bibinfo {author} {\bibfnamefont {C.~W.}\ \bibnamefont {Hsu}}, \bibinfo {author} {\bibfnamefont {A.}~\bibnamefont {Goetschy}}, \bibinfo {author} {\bibfnamefont {S.}~\bibnamefont {Bittner}}, \bibinfo {author} {\bibfnamefont {S.}~\bibnamefont {Rotter}}, \bibinfo {author} {\bibfnamefont {A.}~\bibnamefont {Yamilov}},\ and\ \bibinfo {author} {\bibfnamefont {H.}~\bibnamefont {Cao}},\ }\bibfield  {title} {\bibinfo {title} {Angular memory effect of transmission eigenchannels},\ }\href {https://doi.org/10.1103/PhysRevLett.123.203901} {\bibfield  {journal} {\bibinfo  {journal} {Phys. Rev. Lett.}\ }\textbf {\bibinfo {volume} {123}},\ \bibinfo {pages} {203901} (\bibinfo {year} {2019})}\BibitemShut {NoStop}%
\bibitem [{\citenamefont {Vellekoop}\ and\ \citenamefont {Aegerter}(2010)}]{2010_Vellekoop_OL}%
  \BibitemOpen
  \bibfield  {author} {\bibinfo {author} {\bibfnamefont {I.~M.}\ \bibnamefont {Vellekoop}}\ and\ \bibinfo {author} {\bibfnamefont {C.}~\bibnamefont {Aegerter}},\ }\bibfield  {title} {\bibinfo {title} {Scattered light fluorescence microscopy: imaging through turbid layers},\ }\href@noop {} {\bibfield  {journal} {\bibinfo  {journal} {Opt. Lett.}\ }\textbf {\bibinfo {volume} {35}},\ \bibinfo {pages} {1245} (\bibinfo {year} {2010})}\BibitemShut {NoStop}%
\bibitem [{\citenamefont {Hsieh}\ \emph {et~al.}(2010)\citenamefont {Hsieh}, \citenamefont {Pu}, \citenamefont {Grange}, \citenamefont {Laporte},\ and\ \citenamefont {Psaltis}}]{2010_Psaltis_OE}%
  \BibitemOpen
  \bibfield  {author} {\bibinfo {author} {\bibfnamefont {C.~L.}\ \bibnamefont {Hsieh}}, \bibinfo {author} {\bibfnamefont {Y.}~\bibnamefont {Pu}}, \bibinfo {author} {\bibfnamefont {R.}~\bibnamefont {Grange}}, \bibinfo {author} {\bibfnamefont {G.}~\bibnamefont {Laporte}},\ and\ \bibinfo {author} {\bibfnamefont {D.}~\bibnamefont {Psaltis}},\ }\bibfield  {title} {\bibinfo {title} {Imaging through turbid layers by scanning the phase conjugated second harmonic radiation from a nanoparticle},\ }\href@noop {} {\bibfield  {journal} {\bibinfo  {journal} {Opt. Express}\ }\textbf {\bibinfo {volume} {18}},\ \bibinfo {pages} {20723} (\bibinfo {year} {2010})}\BibitemShut {NoStop}%
\bibitem [{\citenamefont {van Putten}\ \emph {et~al.}(2011)\citenamefont {van Putten}, \citenamefont {Akbulut}, \citenamefont {Bertolotti}, \citenamefont {Vos}, \citenamefont {Lagendijk},\ and\ \citenamefont {Mosk}}]{2011_VanPutten_PRL}%
  \BibitemOpen
  \bibfield  {author} {\bibinfo {author} {\bibfnamefont {E.~G.}\ \bibnamefont {van Putten}}, \bibinfo {author} {\bibfnamefont {D.}~\bibnamefont {Akbulut}}, \bibinfo {author} {\bibfnamefont {J.}~\bibnamefont {Bertolotti}}, \bibinfo {author} {\bibfnamefont {W.~L.}\ \bibnamefont {Vos}}, \bibinfo {author} {\bibfnamefont {A.}~\bibnamefont {Lagendijk}},\ and\ \bibinfo {author} {\bibfnamefont {A.~P.}\ \bibnamefont {Mosk}},\ }\bibfield  {title} {\bibinfo {title} {Scattering lens resolves sub-100 nm structures with visible light},\ }\href@noop {} {\bibfield  {journal} {\bibinfo  {journal} {Phys. Rev. Lett.}\ }\textbf {\bibinfo {volume} {106}},\ \bibinfo {pages} {193905} (\bibinfo {year} {2011})}\BibitemShut {NoStop}%
\bibitem [{\citenamefont {Katz}\ \emph {et~al.}(2012)\citenamefont {Katz}, \citenamefont {Small},\ and\ \citenamefont {Silberberg}}]{2012_Katz_NatPhoton}%
  \BibitemOpen
  \bibfield  {author} {\bibinfo {author} {\bibfnamefont {O.}~\bibnamefont {Katz}}, \bibinfo {author} {\bibfnamefont {E.}~\bibnamefont {Small}},\ and\ \bibinfo {author} {\bibfnamefont {Y.}~\bibnamefont {Silberberg}},\ }\bibfield  {title} {\bibinfo {title} {Looking around corners and through thin turbid layers in real time with scattered incoherent light},\ }\href@noop {} {\bibfield  {journal} {\bibinfo  {journal} {Nat. Photonics}\ }\textbf {\bibinfo {volume} {6}},\ \bibinfo {pages} {549} (\bibinfo {year} {2012})}\BibitemShut {NoStop}%
\bibitem [{\citenamefont {Goorden}\ \emph {et~al.}(2014)\citenamefont {Goorden}, \citenamefont {Horstmann}, \citenamefont {Mosk}, \citenamefont {{\v{S}}kori{\'c}},\ and\ \citenamefont {Pinkse}}]{goorden2014quantum}%
  \BibitemOpen
  \bibfield  {author} {\bibinfo {author} {\bibfnamefont {S.~A.}\ \bibnamefont {Goorden}}, \bibinfo {author} {\bibfnamefont {M.}~\bibnamefont {Horstmann}}, \bibinfo {author} {\bibfnamefont {A.~P.}\ \bibnamefont {Mosk}}, \bibinfo {author} {\bibfnamefont {B.}~\bibnamefont {{\v{S}}kori{\'c}}},\ and\ \bibinfo {author} {\bibfnamefont {P.~W.}\ \bibnamefont {Pinkse}},\ }\bibfield  {title} {\bibinfo {title} {Quantum-secure authentication of a physical unclonable key},\ }\href@noop {} {\bibfield  {journal} {\bibinfo  {journal} {Optica}\ }\textbf {\bibinfo {volume} {1}},\ \bibinfo {pages} {421} (\bibinfo {year} {2014})}\BibitemShut {NoStop}%
\bibitem [{\citenamefont {Huisman}\ \emph {et~al.}(2014)\citenamefont {Huisman}, \citenamefont {Huisman}, \citenamefont {Goorden}, \citenamefont {Mosk},\ and\ \citenamefont {Pinkse}}]{Huisman2014}%
  \BibitemOpen
  \bibfield  {author} {\bibinfo {author} {\bibfnamefont {S.~R.}\ \bibnamefont {Huisman}}, \bibinfo {author} {\bibfnamefont {T.~J.}\ \bibnamefont {Huisman}}, \bibinfo {author} {\bibfnamefont {S.~A.}\ \bibnamefont {Goorden}}, \bibinfo {author} {\bibfnamefont {A.~P.}\ \bibnamefont {Mosk}},\ and\ \bibinfo {author} {\bibfnamefont {P.~W.~H.}\ \bibnamefont {Pinkse}},\ }\bibfield  {title} {\bibinfo {title} {Programming balanced optical beam splitters in white paint},\ }\href {https://doi.org/10.1364/oe.22.008320} {\bibfield  {journal} {\bibinfo  {journal} {Opt. Express}\ }\textbf {\bibinfo {volume} {22}},\ \bibinfo {pages} {8320} (\bibinfo {year} {2014})}\BibitemShut {NoStop}%
\bibitem [{\citenamefont {Huisman}\ \emph {et~al.}(2015)\citenamefont {Huisman}, \citenamefont {Huisman}, \citenamefont {Wolterink}, \citenamefont {Mosk},\ and\ \citenamefont {Pinkse}}]{Huisman2015}%
  \BibitemOpen
  \bibfield  {author} {\bibinfo {author} {\bibfnamefont {S.~R.}\ \bibnamefont {Huisman}}, \bibinfo {author} {\bibfnamefont {T.~J.}\ \bibnamefont {Huisman}}, \bibinfo {author} {\bibfnamefont {T.~A.~W.}\ \bibnamefont {Wolterink}}, \bibinfo {author} {\bibfnamefont {A.~P.}\ \bibnamefont {Mosk}},\ and\ \bibinfo {author} {\bibfnamefont {P.~W.~H.}\ \bibnamefont {Pinkse}},\ }\bibfield  {title} {\bibinfo {title} {Programmable multiport optical circuits in opaque scattering materials},\ }\href {https://doi.org/10.1364/oe.23.003102} {\bibfield  {journal} {\bibinfo  {journal} {Opt. Express}\ }\textbf {\bibinfo {volume} {23}},\ \bibinfo {pages} {3102} (\bibinfo {year} {2015})}\BibitemShut {NoStop}%
\bibitem [{\citenamefont {Wolterink}\ \emph {et~al.}(2016)\citenamefont {Wolterink}, \citenamefont {Uppu}, \citenamefont {Ctistis}, \citenamefont {Vos}, \citenamefont {Boller},\ and\ \citenamefont {Pinkse}}]{wolterink2016programmable}%
  \BibitemOpen
  \bibfield  {author} {\bibinfo {author} {\bibfnamefont {T.~A.}\ \bibnamefont {Wolterink}}, \bibinfo {author} {\bibfnamefont {R.}~\bibnamefont {Uppu}}, \bibinfo {author} {\bibfnamefont {G.}~\bibnamefont {Ctistis}}, \bibinfo {author} {\bibfnamefont {W.~L.}\ \bibnamefont {Vos}}, \bibinfo {author} {\bibfnamefont {K.-J.}\ \bibnamefont {Boller}},\ and\ \bibinfo {author} {\bibfnamefont {P.~W.}\ \bibnamefont {Pinkse}},\ }\bibfield  {title} {\bibinfo {title} {Programmable two-photon quantum interference in $10^3$ channels in opaque scattering media},\ }\href@noop {} {\bibfield  {journal} {\bibinfo  {journal} {Phys. Rev. A}\ }\textbf {\bibinfo {volume} {93}},\ \bibinfo {pages} {053817} (\bibinfo {year} {2016})}\BibitemShut {NoStop}%
\bibitem [{\citenamefont {Peng}\ \emph {et~al.}(2019)\citenamefont {Peng}, \citenamefont {Li}, \citenamefont {An}, \citenamefont {Yu}, \citenamefont {Zhou}, \citenamefont {Bai}, \citenamefont {Liang}, \citenamefont {Lei}, \citenamefont {Zhang}, \citenamefont {Yao},\ and\ \citenamefont {Zhang}}]{Peng2019}%
  \BibitemOpen
  \bibfield  {author} {\bibinfo {author} {\bibfnamefont {T.}~\bibnamefont {Peng}}, \bibinfo {author} {\bibfnamefont {R.}~\bibnamefont {Li}}, \bibinfo {author} {\bibfnamefont {S.}~\bibnamefont {An}}, \bibinfo {author} {\bibfnamefont {X.}~\bibnamefont {Yu}}, \bibinfo {author} {\bibfnamefont {M.}~\bibnamefont {Zhou}}, \bibinfo {author} {\bibfnamefont {C.}~\bibnamefont {Bai}}, \bibinfo {author} {\bibfnamefont {Y.}~\bibnamefont {Liang}}, \bibinfo {author} {\bibfnamefont {M.}~\bibnamefont {Lei}}, \bibinfo {author} {\bibfnamefont {C.}~\bibnamefont {Zhang}}, \bibinfo {author} {\bibfnamefont {B.}~\bibnamefont {Yao}},\ and\ \bibinfo {author} {\bibfnamefont {P.}~\bibnamefont {Zhang}},\ }\bibfield  {title} {\bibinfo {title} {Real-time optical manipulation of particles through turbid media},\ }\href {https://doi.org/10.1364/oe.27.004858} {\bibfield  {journal} {\bibinfo  {journal} {Opt. Express}\ }\textbf {\bibinfo {volume} {27}},\ \bibinfo {pages} {4858} (\bibinfo {year} {2019})}\BibitemShut {NoStop}%
\bibitem [{\citenamefont {Wang}\ \emph {et~al.}(2015)\citenamefont {Wang}, \citenamefont {He}, \citenamefont {Zhuang}, \citenamefont {Xie}, \citenamefont {Yang}, \citenamefont {Cai}, \citenamefont {Gu},\ and\ \citenamefont {Zhou}}]{wang2015controlled}%
  \BibitemOpen
  \bibfield  {author} {\bibinfo {author} {\bibfnamefont {F.}~\bibnamefont {Wang}}, \bibinfo {author} {\bibfnamefont {H.}~\bibnamefont {He}}, \bibinfo {author} {\bibfnamefont {H.}~\bibnamefont {Zhuang}}, \bibinfo {author} {\bibfnamefont {X.}~\bibnamefont {Xie}}, \bibinfo {author} {\bibfnamefont {Z.}~\bibnamefont {Yang}}, \bibinfo {author} {\bibfnamefont {Z.}~\bibnamefont {Cai}}, \bibinfo {author} {\bibfnamefont {H.}~\bibnamefont {Gu}},\ and\ \bibinfo {author} {\bibfnamefont {J.}~\bibnamefont {Zhou}},\ }\bibfield  {title} {\bibinfo {title} {Controlled light field concentration through turbid biological membrane for phototherapy},\ }\href@noop {} {\bibfield  {journal} {\bibinfo  {journal} {Biomed. Opt. Express}\ }\textbf {\bibinfo {volume} {6}},\ \bibinfo {pages} {2237} (\bibinfo {year} {2015})}\BibitemShut {NoStop}%
\bibitem [{\citenamefont {Yanik}\ \emph {et~al.}(2004)\citenamefont {Yanik}, \citenamefont {Cinar}, \citenamefont {Cinar}, \citenamefont {Chisholm}, \citenamefont {Jin},\ and\ \citenamefont {Ben-Yakar}}]{yanik2004functional}%
  \BibitemOpen
  \bibfield  {author} {\bibinfo {author} {\bibfnamefont {M.~F.}\ \bibnamefont {Yanik}}, \bibinfo {author} {\bibfnamefont {H.}~\bibnamefont {Cinar}}, \bibinfo {author} {\bibfnamefont {H.~N.}\ \bibnamefont {Cinar}}, \bibinfo {author} {\bibfnamefont {A.~D.}\ \bibnamefont {Chisholm}}, \bibinfo {author} {\bibfnamefont {Y.}~\bibnamefont {Jin}},\ and\ \bibinfo {author} {\bibfnamefont {A.}~\bibnamefont {Ben-Yakar}},\ }\bibfield  {title} {\bibinfo {title} {Functional regeneration after laser axotomy},\ }\href@noop {} {\bibfield  {journal} {\bibinfo  {journal} {Nature}\ }\textbf {\bibinfo {volume} {432}},\ \bibinfo {pages} {822} (\bibinfo {year} {2004})}\BibitemShut {NoStop}%
\bibitem [{\citenamefont {Fenno}\ \emph {et~al.}(2011)\citenamefont {Fenno}, \citenamefont {Yizhar},\ and\ \citenamefont {Deisseroth}}]{fenno2011development}%
  \BibitemOpen
  \bibfield  {author} {\bibinfo {author} {\bibfnamefont {L.}~\bibnamefont {Fenno}}, \bibinfo {author} {\bibfnamefont {O.}~\bibnamefont {Yizhar}},\ and\ \bibinfo {author} {\bibfnamefont {K.}~\bibnamefont {Deisseroth}},\ }\bibfield  {title} {\bibinfo {title} {The development and application of optogenetics},\ }\href@noop {} {\bibfield  {journal} {\bibinfo  {journal} {Annu. Rev. Neurosci.}\ }\textbf {\bibinfo {volume} {34}},\ \bibinfo {pages} {389} (\bibinfo {year} {2011})}\BibitemShut {NoStop}%
\bibitem [{\citenamefont {Ruan}\ \emph {et~al.}(2017)\citenamefont {Ruan}, \citenamefont {Brake}, \citenamefont {Robinson}, \citenamefont {Liu}, \citenamefont {Jang}, \citenamefont {Xiao}, \citenamefont {Zhou}, \citenamefont {Gradinaru},\ and\ \citenamefont {Yang}}]{ruan2017deep}%
  \BibitemOpen
  \bibfield  {author} {\bibinfo {author} {\bibfnamefont {H.}~\bibnamefont {Ruan}}, \bibinfo {author} {\bibfnamefont {J.}~\bibnamefont {Brake}}, \bibinfo {author} {\bibfnamefont {J.~E.}\ \bibnamefont {Robinson}}, \bibinfo {author} {\bibfnamefont {Y.}~\bibnamefont {Liu}}, \bibinfo {author} {\bibfnamefont {M.}~\bibnamefont {Jang}}, \bibinfo {author} {\bibfnamefont {C.}~\bibnamefont {Xiao}}, \bibinfo {author} {\bibfnamefont {C.}~\bibnamefont {Zhou}}, \bibinfo {author} {\bibfnamefont {V.}~\bibnamefont {Gradinaru}},\ and\ \bibinfo {author} {\bibfnamefont {C.}~\bibnamefont {Yang}},\ }\bibfield  {title} {\bibinfo {title} {Deep tissue optical focusing and optogenetic modulation with time-reversed ultrasonically encoded light},\ }\href@noop {} {\bibfield  {journal} {\bibinfo  {journal} {Sci. Adv.}\ }\textbf {\bibinfo {volume} {3}},\ \bibinfo {pages} {eaao5520} (\bibinfo {year} {2017})}\BibitemShut {NoStop}%
\bibitem [{\citenamefont {P{\'e}gard}\ \emph {et~al.}(2017)\citenamefont {P{\'e}gard}, \citenamefont {Mardinly}, \citenamefont {Oldenburg}, \citenamefont {Sridharan}, \citenamefont {Waller},\ and\ \citenamefont {Adesnik}}]{pegard2017three}%
  \BibitemOpen
  \bibfield  {author} {\bibinfo {author} {\bibfnamefont {N.~C.}\ \bibnamefont {P{\'e}gard}}, \bibinfo {author} {\bibfnamefont {A.~R.}\ \bibnamefont {Mardinly}}, \bibinfo {author} {\bibfnamefont {I.~A.}\ \bibnamefont {Oldenburg}}, \bibinfo {author} {\bibfnamefont {S.}~\bibnamefont {Sridharan}}, \bibinfo {author} {\bibfnamefont {L.}~\bibnamefont {Waller}},\ and\ \bibinfo {author} {\bibfnamefont {H.}~\bibnamefont {Adesnik}},\ }\bibfield  {title} {\bibinfo {title} {Three-dimensional scanless holographic optogenetics with temporal focusing ({3D-SHOT})},\ }\href@noop {} {\bibfield  {journal} {\bibinfo  {journal} {Nat. Commun.}\ }\textbf {\bibinfo {volume} {8}},\ \bibinfo {pages} {1228} (\bibinfo {year} {2017})}\BibitemShut {NoStop}%
\bibitem [{\citenamefont {Popoff}\ \emph {et~al.}(2010{\natexlab{a}})\citenamefont {Popoff}, \citenamefont {Lerosey}, \citenamefont {Carminati}, \citenamefont {Fink}, \citenamefont {Boccara},\ and\ \citenamefont {Gigan}}]{2010_Popoff_PRL}%
  \BibitemOpen
  \bibfield  {author} {\bibinfo {author} {\bibfnamefont {S.~M.}\ \bibnamefont {Popoff}}, \bibinfo {author} {\bibfnamefont {G.}~\bibnamefont {Lerosey}}, \bibinfo {author} {\bibfnamefont {R.}~\bibnamefont {Carminati}}, \bibinfo {author} {\bibfnamefont {M.}~\bibnamefont {Fink}}, \bibinfo {author} {\bibfnamefont {A.~C.}\ \bibnamefont {Boccara}},\ and\ \bibinfo {author} {\bibfnamefont {S.}~\bibnamefont {Gigan}},\ }\bibfield  {title} {\bibinfo {title} {Measuring the transmission matrix in optics: an approach to the study and control of light propagation in disordered media},\ }\href@noop {} {\bibfield  {journal} {\bibinfo  {journal} {Phys. Rev. Lett.}\ }\textbf {\bibinfo {volume} {104}},\ \bibinfo {pages} {100601} (\bibinfo {year} {2010}{\natexlab{a}})}\BibitemShut {NoStop}%
\bibitem [{\citenamefont {Prada}\ and\ \citenamefont {Fink}(1994)}]{prada1994eigenmodes}%
  \BibitemOpen
  \bibfield  {author} {\bibinfo {author} {\bibfnamefont {C.}~\bibnamefont {Prada}}\ and\ \bibinfo {author} {\bibfnamefont {M.}~\bibnamefont {Fink}},\ }\bibfield  {title} {\bibinfo {title} {Eigenmodes of the time reversal operator: A solution to selective focusing in multiple-target media},\ }\href@noop {} {\bibfield  {journal} {\bibinfo  {journal} {Wave Motion}\ }\textbf {\bibinfo {volume} {20}},\ \bibinfo {pages} {151} (\bibinfo {year} {1994})}\BibitemShut {NoStop}%
\bibitem [{\citenamefont {Vellekoop}\ \emph {et~al.}(2008)\citenamefont {Vellekoop}, \citenamefont {van Putten}, \citenamefont {Lagendijk},\ and\ \citenamefont {Mosk}}]{2008_Vellekoop_OE}%
  \BibitemOpen
  \bibfield  {author} {\bibinfo {author} {\bibfnamefont {I.~M.}\ \bibnamefont {Vellekoop}}, \bibinfo {author} {\bibfnamefont {E.~G.}\ \bibnamefont {van Putten}}, \bibinfo {author} {\bibfnamefont {A.}~\bibnamefont {Lagendijk}},\ and\ \bibinfo {author} {\bibfnamefont {A.~P.}\ \bibnamefont {Mosk}},\ }\bibfield  {title} {\bibinfo {title} {Demixing light paths inside disordered metamaterials},\ }\href@noop {} {\bibfield  {journal} {\bibinfo  {journal} {Opt. Express}\ }\textbf {\bibinfo {volume} {16}},\ \bibinfo {pages} {67} (\bibinfo {year} {2008})}\BibitemShut {NoStop}%
\bibitem [{\citenamefont {van Putten}\ \emph {et~al.}(2012)\citenamefont {van Putten}, \citenamefont {Lagendijk},\ and\ \citenamefont {Mosk}}]{van2012nonimaging}%
  \BibitemOpen
  \bibfield  {author} {\bibinfo {author} {\bibfnamefont {E.}~\bibnamefont {van Putten}}, \bibinfo {author} {\bibfnamefont {A.}~\bibnamefont {Lagendijk}},\ and\ \bibinfo {author} {\bibfnamefont {A.}~\bibnamefont {Mosk}},\ }\bibfield  {title} {\bibinfo {title} {Nonimaging speckle interferometry for high-speed nanometer-scale position detection},\ }\href@noop {} {\bibfield  {journal} {\bibinfo  {journal} {Opt. Lett.}\ }\textbf {\bibinfo {volume} {37}},\ \bibinfo {pages} {1070} (\bibinfo {year} {2012})}\BibitemShut {NoStop}%
\bibitem [{\citenamefont {Xu}\ \emph {et~al.}(2017)\citenamefont {Xu}, \citenamefont {Ruan}, \citenamefont {Liu}, \citenamefont {Zhou},\ and\ \citenamefont {Yang}}]{xu2017focusing}%
  \BibitemOpen
  \bibfield  {author} {\bibinfo {author} {\bibfnamefont {J.}~\bibnamefont {Xu}}, \bibinfo {author} {\bibfnamefont {H.}~\bibnamefont {Ruan}}, \bibinfo {author} {\bibfnamefont {Y.}~\bibnamefont {Liu}}, \bibinfo {author} {\bibfnamefont {H.}~\bibnamefont {Zhou}},\ and\ \bibinfo {author} {\bibfnamefont {C.}~\bibnamefont {Yang}},\ }\bibfield  {title} {\bibinfo {title} {Focusing light through scattering media by transmission matrix inversion},\ }\href@noop {} {\bibfield  {journal} {\bibinfo  {journal} {Opt. Express}\ }\textbf {\bibinfo {volume} {25}},\ \bibinfo {pages} {27234} (\bibinfo {year} {2017})}\BibitemShut {NoStop}%
\bibitem [{\citenamefont {Feng}\ \emph {et~al.}(2019)\citenamefont {Feng}, \citenamefont {Yang}, \citenamefont {Xu}, \citenamefont {Zhang}, \citenamefont {Ding},\ and\ \citenamefont {Liu}}]{Feng2019}%
  \BibitemOpen
  \bibfield  {author} {\bibinfo {author} {\bibfnamefont {Q.}~\bibnamefont {Feng}}, \bibinfo {author} {\bibfnamefont {F.}~\bibnamefont {Yang}}, \bibinfo {author} {\bibfnamefont {X.}~\bibnamefont {Xu}}, \bibinfo {author} {\bibfnamefont {B.}~\bibnamefont {Zhang}}, \bibinfo {author} {\bibfnamefont {Y.}~\bibnamefont {Ding}},\ and\ \bibinfo {author} {\bibfnamefont {Q.}~\bibnamefont {Liu}},\ }\bibfield  {title} {\bibinfo {title} {Multi-objective optimization genetic algorithm for multi-point light focusing in wavefront shaping},\ }\href {https://doi.org/10.1364/oe.27.036459} {\bibfield  {journal} {\bibinfo  {journal} {Opt. Express}\ }\textbf {\bibinfo {volume} {27}},\ \bibinfo {pages} {36459} (\bibinfo {year} {2019})}\BibitemShut {NoStop}%
\bibitem [{\citenamefont {Hu}\ \emph {et~al.}(2023)\citenamefont {Hu}, \citenamefont {Hu}, \citenamefont {Zhou},\ and\ \citenamefont {Ding}}]{Hu2023}%
  \BibitemOpen
  \bibfield  {author} {\bibinfo {author} {\bibfnamefont {Y.}~\bibnamefont {Hu}}, \bibinfo {author} {\bibfnamefont {M.}~\bibnamefont {Hu}}, \bibinfo {author} {\bibfnamefont {J.}~\bibnamefont {Zhou}},\ and\ \bibinfo {author} {\bibfnamefont {Y.}~\bibnamefont {Ding}},\ }\bibfield  {title} {\bibinfo {title} {Multi-point optical focusing based on enhanced multi-objective optimized wavefront shaping},\ }\href {https://doi.org/10.1016/j.optcom.2023.129836} {\bibfield  {journal} {\bibinfo  {journal} {Opt. Commun.}\ }\textbf {\bibinfo {volume} {547}},\ \bibinfo {pages} {129836} (\bibinfo {year} {2023})}\BibitemShut {NoStop}%
\bibitem [{\citenamefont {Hong}\ \emph {et~al.}(2024)\citenamefont {Hong}, \citenamefont {Liang},\ and\ \citenamefont {Zhang}}]{hong2024robust}%
  \BibitemOpen
  \bibfield  {author} {\bibinfo {author} {\bibfnamefont {P.}~\bibnamefont {Hong}}, \bibinfo {author} {\bibfnamefont {Y.}~\bibnamefont {Liang}},\ and\ \bibinfo {author} {\bibfnamefont {G.}~\bibnamefont {Zhang}},\ }\bibfield  {title} {\bibinfo {title} {Robust multiple focusing through scattering media via feedback wavefront shaping},\ }\href@noop {} {\bibfield  {journal} {\bibinfo  {journal} {Opt. Laser Technol.}\ }\textbf {\bibinfo {volume} {176}},\ \bibinfo {pages} {110939} (\bibinfo {year} {2024})}\BibitemShut {NoStop}%
\bibitem [{\citenamefont {Cheng}\ and\ \citenamefont {Chen}(2025)}]{cheng2025binary}%
  \BibitemOpen
  \bibfield  {author} {\bibinfo {author} {\bibfnamefont {Z.}~\bibnamefont {Cheng}}\ and\ \bibinfo {author} {\bibfnamefont {Y.}~\bibnamefont {Chen}},\ }\bibfield  {title} {\bibinfo {title} {Binary-intensity-synthetic digital optical phase conjugation for multifocal control through scattering media},\ }\href@noop {} {\bibfield  {journal} {\bibinfo  {journal} {Opt. Lett.}\ }\textbf {\bibinfo {volume} {50}},\ \bibinfo {pages} {5530} (\bibinfo {year} {2025})}\BibitemShut {NoStop}%
\bibitem [{\citenamefont {Popoff}\ \emph {et~al.}(2010{\natexlab{b}})\citenamefont {Popoff}, \citenamefont {Lerosey}, \citenamefont {Fink}, \citenamefont {Boccara},\ and\ \citenamefont {Gigan}}]{popoff2010image}%
  \BibitemOpen
  \bibfield  {author} {\bibinfo {author} {\bibfnamefont {S.}~\bibnamefont {Popoff}}, \bibinfo {author} {\bibfnamefont {G.}~\bibnamefont {Lerosey}}, \bibinfo {author} {\bibfnamefont {M.}~\bibnamefont {Fink}}, \bibinfo {author} {\bibfnamefont {A.~C.}\ \bibnamefont {Boccara}},\ and\ \bibinfo {author} {\bibfnamefont {S.}~\bibnamefont {Gigan}},\ }\bibfield  {title} {\bibinfo {title} {Image transmission through an opaque material},\ }\href@noop {} {\bibfield  {journal} {\bibinfo  {journal} {Nat. Commun.}\ }\textbf {\bibinfo {volume} {1}},\ \bibinfo {pages} {81} (\bibinfo {year} {2010}{\natexlab{b}})}\BibitemShut {NoStop}%
\bibitem [{\citenamefont {Zhou}\ \emph {et~al.}(2014)\citenamefont {Zhou}, \citenamefont {Ruan}, \citenamefont {Yang},\ and\ \citenamefont {Judkewitz}}]{Zhou2014}%
  \BibitemOpen
  \bibfield  {author} {\bibinfo {author} {\bibfnamefont {E.~H.}\ \bibnamefont {Zhou}}, \bibinfo {author} {\bibfnamefont {H.}~\bibnamefont {Ruan}}, \bibinfo {author} {\bibfnamefont {C.}~\bibnamefont {Yang}},\ and\ \bibinfo {author} {\bibfnamefont {B.}~\bibnamefont {Judkewitz}},\ }\bibfield  {title} {\bibinfo {title} {Focusing on moving targets through scattering samples},\ }\href {https://doi.org/10.1364/optica.1.000227} {\bibfield  {journal} {\bibinfo  {journal} {Optica}\ }\textbf {\bibinfo {volume} {1}},\ \bibinfo {pages} {227} (\bibinfo {year} {2014})}\BibitemShut {NoStop}%
\bibitem [{\citenamefont {Vellekoop}\ and\ \citenamefont {Mosk}(2008{\natexlab{b}})}]{vellekoop2008phase}%
  \BibitemOpen
  \bibfield  {author} {\bibinfo {author} {\bibfnamefont {I.~M.}\ \bibnamefont {Vellekoop}}\ and\ \bibinfo {author} {\bibfnamefont {A.}~\bibnamefont {Mosk}},\ }\bibfield  {title} {\bibinfo {title} {Phase control algorithms for focusing light through turbid media},\ }\href@noop {} {\bibfield  {journal} {\bibinfo  {journal} {Opt. Commun.}\ }\textbf {\bibinfo {volume} {281}},\ \bibinfo {pages} {3071} (\bibinfo {year} {2008}{\natexlab{b}})}\BibitemShut {NoStop}%
\bibitem [{\citenamefont {Beenakker}(1997)}]{beenakker1997random}%
  \BibitemOpen
  \bibfield  {author} {\bibinfo {author} {\bibfnamefont {C.~W.}\ \bibnamefont {Beenakker}},\ }\bibfield  {title} {\bibinfo {title} {Random-matrix theory of quantum transport},\ }\href@noop {} {\bibfield  {journal} {\bibinfo  {journal} {Rev. Mod. Phys.}\ }\textbf {\bibinfo {volume} {69}},\ \bibinfo {pages} {731} (\bibinfo {year} {1997})}\BibitemShut {NoStop}%
\bibitem [{\citenamefont {Mar{\v{c}}enko}\ and\ \citenamefont {Pastur}(1967)}]{marvcenko1967distribution}%
  \BibitemOpen
  \bibfield  {author} {\bibinfo {author} {\bibfnamefont {V.~A.}\ \bibnamefont {Mar{\v{c}}enko}}\ and\ \bibinfo {author} {\bibfnamefont {L.~A.}\ \bibnamefont {Pastur}},\ }\bibfield  {title} {\bibinfo {title} {Distribution of eigenvalues for some sets of random matrices},\ }\href@noop {} {\bibfield  {journal} {\bibinfo  {journal} {Math. USSR Sbornik}\ }\textbf {\bibinfo {volume} {1}},\ \bibinfo {pages} {457} (\bibinfo {year} {1967})}\BibitemShut {NoStop}%
\bibitem [{\citenamefont {Mehta}(2004)}]{mehta2004random}%
  \BibitemOpen
  \bibfield  {author} {\bibinfo {author} {\bibfnamefont {M.~L.}\ \bibnamefont {Mehta}},\ }\href@noop {} {\emph {\bibinfo {title} {Random matrices}}},\ Vol.\ \bibinfo {volume} {142}\ (\bibinfo  {publisher} {Elsevier},\ \bibinfo {year} {2004})\BibitemShut {NoStop}%
\bibitem [{\citenamefont {Forrester}(2010)}]{forrester2010log}%
  \BibitemOpen
  \bibfield  {author} {\bibinfo {author} {\bibfnamefont {P.~J.}\ \bibnamefont {Forrester}},\ }\href@noop {} {\emph {\bibinfo {title} {Log-gases and random matrices (LMS-34)}}}\ (\bibinfo  {publisher} {Princeton University Press},\ \bibinfo {year} {2010})\BibitemShut {NoStop}%
\bibitem [{\citenamefont {Paniagua-Diaz}\ \emph {et~al.}(2023)\citenamefont {Paniagua-Diaz}, \citenamefont {Barnes},\ and\ \citenamefont {Bertolotti}}]{PaniaguaDiaz2023}%
  \BibitemOpen
  \bibfield  {author} {\bibinfo {author} {\bibfnamefont {A.~M.}\ \bibnamefont {Paniagua-Diaz}}, \bibinfo {author} {\bibfnamefont {W.~L.}\ \bibnamefont {Barnes}},\ and\ \bibinfo {author} {\bibfnamefont {J.}~\bibnamefont {Bertolotti}},\ }\bibfield  {title} {\bibinfo {title} {Wavefront shaping to improve beam quality: converting a speckle pattern into a {G}aussian spot},\ }\href {https://doi.org/10.1088/1402-4896/acbbab} {\bibfield  {journal} {\bibinfo  {journal} {Phys. Scr.}\ }\textbf {\bibinfo {volume} {98}},\ \bibinfo {pages} {035516} (\bibinfo {year} {2023})}\BibitemShut {NoStop}%
\bibitem [{\citenamefont {Ojambati}\ \emph {et~al.}(2016{\natexlab{b}})\citenamefont {Ojambati}, \citenamefont {Hosmer-Quint}, \citenamefont {Gorter}, \citenamefont {Mosk},\ and\ \citenamefont {Vos}}]{ojambati2016controlling}%
  \BibitemOpen
  \bibfield  {author} {\bibinfo {author} {\bibfnamefont {O.~S.}\ \bibnamefont {Ojambati}}, \bibinfo {author} {\bibfnamefont {J.~T.}\ \bibnamefont {Hosmer-Quint}}, \bibinfo {author} {\bibfnamefont {K.-J.}\ \bibnamefont {Gorter}}, \bibinfo {author} {\bibfnamefont {A.~P.}\ \bibnamefont {Mosk}},\ and\ \bibinfo {author} {\bibfnamefont {W.~L.}\ \bibnamefont {Vos}},\ }\bibfield  {title} {\bibinfo {title} {Controlling the intensity of light in large areas at the interfaces of a scattering medium},\ }\href@noop {} {\bibfield  {journal} {\bibinfo  {journal} {Phys. Rev. A}\ }\textbf {\bibinfo {volume} {94}},\ \bibinfo {pages} {043834} (\bibinfo {year} {2016}{\natexlab{b}})}\BibitemShut {NoStop}%
\bibitem [{\citenamefont {Shaughnessy}\ \emph {et~al.}(2024)\citenamefont {Shaughnessy}, \citenamefont {McIntosh}, \citenamefont {Goetschy}, \citenamefont {Hsu}, \citenamefont {Bender}, \citenamefont {Y{\i}lmaz}, \citenamefont {Yamilov},\ and\ \citenamefont {Cao}}]{shaughnessy2024multiregion}%
  \BibitemOpen
  \bibfield  {author} {\bibinfo {author} {\bibfnamefont {L.}~\bibnamefont {Shaughnessy}}, \bibinfo {author} {\bibfnamefont {R.~E.}\ \bibnamefont {McIntosh}}, \bibinfo {author} {\bibfnamefont {A.}~\bibnamefont {Goetschy}}, \bibinfo {author} {\bibfnamefont {C.~W.}\ \bibnamefont {Hsu}}, \bibinfo {author} {\bibfnamefont {N.}~\bibnamefont {Bender}}, \bibinfo {author} {\bibfnamefont {H.}~\bibnamefont {Y{\i}lmaz}}, \bibinfo {author} {\bibfnamefont {A.}~\bibnamefont {Yamilov}},\ and\ \bibinfo {author} {\bibfnamefont {H.}~\bibnamefont {Cao}},\ }\bibfield  {title} {\bibinfo {title} {Multiregion light control in diffusive media via wavefront shaping},\ }\href@noop {} {\bibfield  {journal} {\bibinfo  {journal} {Phys. Rev. Lett.}\ }\textbf {\bibinfo {volume} {133}},\ \bibinfo {pages} {146901} (\bibinfo {year} {2024})}\BibitemShut {NoStop}%
\bibitem [{\citenamefont {van Tiggelen}\ \emph {et~al.}(2025)\citenamefont {van Tiggelen}, \citenamefont {Lagendijk},\ and\ \citenamefont {Vos}}]{van2025mesoscopic}%
  \BibitemOpen
  \bibfield  {author} {\bibinfo {author} {\bibfnamefont {B.~A.}\ \bibnamefont {van Tiggelen}}, \bibinfo {author} {\bibfnamefont {A.}~\bibnamefont {Lagendijk}},\ and\ \bibinfo {author} {\bibfnamefont {W.~L.}\ \bibnamefont {Vos}},\ }\bibfield  {title} {\bibinfo {title} {Mesoscopic theory of wavefront shaping to focus waves deep inside disordered media},\ }\href@noop {} {\bibfield  {journal} {\bibinfo  {journal} {Phys. Rev. A}\ }\textbf {\bibinfo {volume} {111}},\ \bibinfo {pages} {053507} (\bibinfo {year} {2025})}\BibitemShut {NoStop}%
\bibitem [{\citenamefont {Akkermans}\ and\ \citenamefont {Montambaux}(2007)}]{2007_Akkermans}%
  \BibitemOpen
  \bibfield  {author} {\bibinfo {author} {\bibfnamefont {E.}~\bibnamefont {Akkermans}}\ and\ \bibinfo {author} {\bibfnamefont {G.}~\bibnamefont {Montambaux}},\ }\href@noop {} {\emph {\bibinfo {title} {Mesoscopic Physics of Electrons and Photons}}}\ (\bibinfo  {publisher} {Cambridge Univ. Press},\ \bibinfo {year} {2007})\BibitemShut {NoStop}%
\bibitem [{\citenamefont {Berkovits}\ and\ \citenamefont {Feng}(1994)}]{berkovits1994correlations}%
  \BibitemOpen
  \bibfield  {author} {\bibinfo {author} {\bibfnamefont {R.}~\bibnamefont {Berkovits}}\ and\ \bibinfo {author} {\bibfnamefont {S.}~\bibnamefont {Feng}},\ }\bibfield  {title} {\bibinfo {title} {Correlations in coherent multiple scattering},\ }\href@noop {} {\bibfield  {journal} {\bibinfo  {journal} {Phys. Rep.}\ }\textbf {\bibinfo {volume} {238}},\ \bibinfo {pages} {135} (\bibinfo {year} {1994})}\BibitemShut {NoStop}%
\bibitem [{\citenamefont {van Rossum}\ and\ \citenamefont {Nieuwenhuizen}(1999)}]{van1999multiple}%
  \BibitemOpen
  \bibfield  {author} {\bibinfo {author} {\bibfnamefont {M.~C.~W.}\ \bibnamefont {van Rossum}}\ and\ \bibinfo {author} {\bibfnamefont {T.~M.}\ \bibnamefont {Nieuwenhuizen}},\ }\bibfield  {title} {\bibinfo {title} {Multiple scattering of classical waves: microscopy, mesoscopy, and diffusion},\ }\href@noop {} {\bibfield  {journal} {\bibinfo  {journal} {Rev. Mod. Phys.}\ }\textbf {\bibinfo {volume} {71}},\ \bibinfo {pages} {313} (\bibinfo {year} {1999})}\BibitemShut {NoStop}%
\bibitem [{\citenamefont {Stephen}\ and\ \citenamefont {Cwilich}(1987)}]{1987_Cwilich}%
  \BibitemOpen
  \bibfield  {author} {\bibinfo {author} {\bibfnamefont {M.~J.}\ \bibnamefont {Stephen}}\ and\ \bibinfo {author} {\bibfnamefont {G.}~\bibnamefont {Cwilich}},\ }\bibfield  {title} {\bibinfo {title} {Intensity correlation functions and fluctuations in light scattered from a random medium},\ }\href@noop {} {\bibfield  {journal} {\bibinfo  {journal} {Phys. Rev. Lett.}\ }\textbf {\bibinfo {volume} {59}},\ \bibinfo {pages} {285} (\bibinfo {year} {1987})}\BibitemShut {NoStop}%
\bibitem [{\citenamefont {Mello}\ \emph {et~al.}(1988)\citenamefont {Mello}, \citenamefont {Akkermans},\ and\ \citenamefont {Shapiro}}]{1988_Mello_Akkermans_Shapiro}%
  \BibitemOpen
  \bibfield  {author} {\bibinfo {author} {\bibfnamefont {P.~A.}\ \bibnamefont {Mello}}, \bibinfo {author} {\bibfnamefont {E.}~\bibnamefont {Akkermans}},\ and\ \bibinfo {author} {\bibfnamefont {B.}~\bibnamefont {Shapiro}},\ }\bibfield  {title} {\bibinfo {title} {Macroscopic approach to correlations in the electronic transmission and reflection from disordered conductors},\ }\href@noop {} {\bibfield  {journal} {\bibinfo  {journal} {Phys. Rev. Lett.}\ }\textbf {\bibinfo {volume} {61}},\ \bibinfo {pages} {459} (\bibinfo {year} {1988})}\BibitemShut {NoStop}%
\bibitem [{\citenamefont {Feng}\ \emph {et~al.}(1988)\citenamefont {Feng}, \citenamefont {Kane}, \citenamefont {Lee},\ and\ \citenamefont {Stone}}]{1988_Stone}%
  \BibitemOpen
  \bibfield  {author} {\bibinfo {author} {\bibfnamefont {S.}~\bibnamefont {Feng}}, \bibinfo {author} {\bibfnamefont {C.}~\bibnamefont {Kane}}, \bibinfo {author} {\bibfnamefont {P.~A.}\ \bibnamefont {Lee}},\ and\ \bibinfo {author} {\bibfnamefont {A.~D.}\ \bibnamefont {Stone}},\ }\bibfield  {title} {\bibinfo {title} {Correlations and fluctuations of coherent wave transmission through disordered media},\ }\href@noop {} {\bibfield  {journal} {\bibinfo  {journal} {Phys. Rev. Lett.}\ }\textbf {\bibinfo {volume} {61}},\ \bibinfo {pages} {834} (\bibinfo {year} {1988})}\BibitemShut {NoStop}%
\bibitem [{\citenamefont {Pnini}\ and\ \citenamefont {Shapiro}(1989)}]{1989_Shapiro}%
  \BibitemOpen
  \bibfield  {author} {\bibinfo {author} {\bibfnamefont {R.}~\bibnamefont {Pnini}}\ and\ \bibinfo {author} {\bibfnamefont {B.}~\bibnamefont {Shapiro}},\ }\bibfield  {title} {\bibinfo {title} {Fluctuations in transmission of waves through disordered slabs},\ }\href@noop {} {\bibfield  {journal} {\bibinfo  {journal} {Phys. Rev. B}\ }\textbf {\bibinfo {volume} {39}},\ \bibinfo {pages} {6986} (\bibinfo {year} {1989})}\BibitemShut {NoStop}%
\bibitem [{\citenamefont {Genack}\ \emph {et~al.}(1990)\citenamefont {Genack}, \citenamefont {Garcia},\ and\ \citenamefont {Polkosnik}}]{1990_Genack}%
  \BibitemOpen
  \bibfield  {author} {\bibinfo {author} {\bibfnamefont {A.~Z.}\ \bibnamefont {Genack}}, \bibinfo {author} {\bibfnamefont {N.}~\bibnamefont {Garcia}},\ and\ \bibinfo {author} {\bibfnamefont {W.}~\bibnamefont {Polkosnik}},\ }\bibfield  {title} {\bibinfo {title} {Long-range intensity correlation in random media},\ }\href@noop {} {\bibfield  {journal} {\bibinfo  {journal} {Phys. Rev. Lett.}\ }\textbf {\bibinfo {volume} {65}},\ \bibinfo {pages} {2129} (\bibinfo {year} {1990})}\BibitemShut {NoStop}%
\bibitem [{\citenamefont {Strudley}\ \emph {et~al.}(2013)\citenamefont {Strudley}, \citenamefont {Zehender}, \citenamefont {Blejean}, \citenamefont {Bakkers},\ and\ \citenamefont {Muskens}}]{2013_Muskens}%
  \BibitemOpen
  \bibfield  {author} {\bibinfo {author} {\bibfnamefont {T.}~\bibnamefont {Strudley}}, \bibinfo {author} {\bibfnamefont {T.}~\bibnamefont {Zehender}}, \bibinfo {author} {\bibfnamefont {C.}~\bibnamefont {Blejean}}, \bibinfo {author} {\bibfnamefont {E.~P. A.~M.}\ \bibnamefont {Bakkers}},\ and\ \bibinfo {author} {\bibfnamefont {O.}~\bibnamefont {Muskens}},\ }\bibfield  {title} {\bibinfo {title} {Mesoscopic light transport by very strong collective multiple scattering in nanowire mats},\ }\href@noop {} {\bibfield  {journal} {\bibinfo  {journal} {Nat. Photonics}\ }\textbf {\bibinfo {volume} {7}},\ \bibinfo {pages} {413} (\bibinfo {year} {2013})}\BibitemShut {NoStop}%
\bibitem [{\citenamefont {Imry}(1986)}]{imry1986active}%
  \BibitemOpen
  \bibfield  {author} {\bibinfo {author} {\bibfnamefont {Y.}~\bibnamefont {Imry}},\ }\bibfield  {title} {\bibinfo {title} {Active transmission channels and universal conductance fluctuations},\ }\href@noop {} {\bibfield  {journal} {\bibinfo  {journal} {Europhys. Lett.}\ }\textbf {\bibinfo {volume} {1}},\ \bibinfo {pages} {249} (\bibinfo {year} {1986})}\BibitemShut {NoStop}%
\bibitem [{\citenamefont {Scheffold}\ and\ \citenamefont {Maret}(1998)}]{scheffold1998universal}%
  \BibitemOpen
  \bibfield  {author} {\bibinfo {author} {\bibfnamefont {F.}~\bibnamefont {Scheffold}}\ and\ \bibinfo {author} {\bibfnamefont {G.}~\bibnamefont {Maret}},\ }\bibfield  {title} {\bibinfo {title} {Universal conductance fluctuations of light},\ }\href@noop {} {\bibfield  {journal} {\bibinfo  {journal} {Phys. Rev. Lett.}\ }\textbf {\bibinfo {volume} {81}},\ \bibinfo {pages} {5800} (\bibinfo {year} {1998})}\BibitemShut {NoStop}%
\bibitem [{\citenamefont {Sarma}\ \emph {et~al.}(2016)\citenamefont {Sarma}, \citenamefont {Yamilov}, \citenamefont {Petrenko}, \citenamefont {Bromberg},\ and\ \citenamefont {Cao}}]{sarma2016control}%
  \BibitemOpen
  \bibfield  {author} {\bibinfo {author} {\bibfnamefont {R.}~\bibnamefont {Sarma}}, \bibinfo {author} {\bibfnamefont {A.~G.}\ \bibnamefont {Yamilov}}, \bibinfo {author} {\bibfnamefont {S.}~\bibnamefont {Petrenko}}, \bibinfo {author} {\bibfnamefont {Y.}~\bibnamefont {Bromberg}},\ and\ \bibinfo {author} {\bibfnamefont {H.}~\bibnamefont {Cao}},\ }\bibfield  {title} {\bibinfo {title} {Control of energy density inside a disordered medium by coupling to open or closed channels},\ }\href@noop {} {\bibfield  {journal} {\bibinfo  {journal} {Phys. Rev. Lett.}\ }\textbf {\bibinfo {volume} {117}},\ \bibinfo {pages} {086803} (\bibinfo {year} {2016})}\BibitemShut {NoStop}%
\bibitem [{\citenamefont {Bender}\ \emph {et~al.}(2020)\citenamefont {Bender}, \citenamefont {Yamilov}, \citenamefont {Y{\i}lmaz},\ and\ \citenamefont {Cao}}]{bender2020fluctuations}%
  \BibitemOpen
  \bibfield  {author} {\bibinfo {author} {\bibfnamefont {N.}~\bibnamefont {Bender}}, \bibinfo {author} {\bibfnamefont {A.}~\bibnamefont {Yamilov}}, \bibinfo {author} {\bibfnamefont {H.}~\bibnamefont {Y{\i}lmaz}},\ and\ \bibinfo {author} {\bibfnamefont {H.}~\bibnamefont {Cao}},\ }\bibfield  {title} {\bibinfo {title} {Fluctuations and correlations of transmission eigenchannels in diffusive media},\ }\href@noop {} {\bibfield  {journal} {\bibinfo  {journal} {Phys. Rev. Lett.}\ }\textbf {\bibinfo {volume} {125}},\ \bibinfo {pages} {165901} (\bibinfo {year} {2020})}\BibitemShut {NoStop}%
\bibitem [{\citenamefont {Lagendijk}\ \emph {et~al.}(2009)\citenamefont {Lagendijk}, \citenamefont {Tiggelen},\ and\ \citenamefont {Wiersma}}]{lagendijk2009fifty}%
  \BibitemOpen
  \bibfield  {author} {\bibinfo {author} {\bibfnamefont {A.}~\bibnamefont {Lagendijk}}, \bibinfo {author} {\bibfnamefont {B.~v.}\ \bibnamefont {Tiggelen}},\ and\ \bibinfo {author} {\bibfnamefont {D.~S.}\ \bibnamefont {Wiersma}},\ }\bibfield  {title} {\bibinfo {title} {Fifty years of {Anderson} localization},\ }\href@noop {} {\bibfield  {journal} {\bibinfo  {journal} {Phys. Today}\ }\textbf {\bibinfo {volume} {62}},\ \bibinfo {pages} {24} (\bibinfo {year} {2009})}\BibitemShut {NoStop}%
\bibitem [{\citenamefont {Hildebrand}\ \emph {et~al.}(2014)\citenamefont {Hildebrand}, \citenamefont {Strybulevych}, \citenamefont {Skipetrov}, \citenamefont {Van~Tiggelen},\ and\ \citenamefont {Page}}]{hildebrand2014observation}%
  \BibitemOpen
  \bibfield  {author} {\bibinfo {author} {\bibfnamefont {W.}~\bibnamefont {Hildebrand}}, \bibinfo {author} {\bibfnamefont {A.}~\bibnamefont {Strybulevych}}, \bibinfo {author} {\bibfnamefont {S.}~\bibnamefont {Skipetrov}}, \bibinfo {author} {\bibfnamefont {B.}~\bibnamefont {Van~Tiggelen}},\ and\ \bibinfo {author} {\bibfnamefont {J.}~\bibnamefont {Page}},\ }\bibfield  {title} {\bibinfo {title} {Observation of infinite-range intensity correlations above, at, and below the mobility edges of the {3D} {Anderson} localization transition},\ }\href@noop {} {\bibfield  {journal} {\bibinfo  {journal} {Phys. Rev. Lett.}\ }\textbf {\bibinfo {volume} {112}},\ \bibinfo {pages} {073902} (\bibinfo {year} {2014})}\BibitemShut {NoStop}%
\bibitem [{\citenamefont {Y{\i}lmaz}\ \emph {et~al.}(2021)\citenamefont {Y{\i}lmaz}, \citenamefont {K{\"u}hmayer}, \citenamefont {Hsu}, \citenamefont {Rotter},\ and\ \citenamefont {Cao}}]{yilmaz2021customizing}%
  \BibitemOpen
  \bibfield  {author} {\bibinfo {author} {\bibfnamefont {H.}~\bibnamefont {Y{\i}lmaz}}, \bibinfo {author} {\bibfnamefont {M.}~\bibnamefont {K{\"u}hmayer}}, \bibinfo {author} {\bibfnamefont {C.~W.}\ \bibnamefont {Hsu}}, \bibinfo {author} {\bibfnamefont {S.}~\bibnamefont {Rotter}},\ and\ \bibinfo {author} {\bibfnamefont {H.}~\bibnamefont {Cao}},\ }\bibfield  {title} {\bibinfo {title} {Customizing the angular memory effect for scattering media},\ }\href@noop {} {\bibfield  {journal} {\bibinfo  {journal} {Phys. Rev. X}\ }\textbf {\bibinfo {volume} {11}},\ \bibinfo {pages} {031010} (\bibinfo {year} {2021})}\BibitemShut {NoStop}%
\bibitem [{\citenamefont {Goetschy}\ and\ \citenamefont {Stone}(2013)}]{2013_Stone_PRL}%
  \BibitemOpen
  \bibfield  {author} {\bibinfo {author} {\bibfnamefont {A.}~\bibnamefont {Goetschy}}\ and\ \bibinfo {author} {\bibfnamefont {A.~D.}\ \bibnamefont {Stone}},\ }\bibfield  {title} {\bibinfo {title} {Filtering random matrices: the effect of incomplete channel control in multiple scattering},\ }\href@noop {} {\bibfield  {journal} {\bibinfo  {journal} {Phys. Rev. Lett.}\ }\textbf {\bibinfo {volume} {111}},\ \bibinfo {pages} {063901} (\bibinfo {year} {2013})}\BibitemShut {NoStop}%
\bibitem [{\citenamefont {Goodman}(2015)}]{goodman2015statistical}%
  \BibitemOpen
  \bibfield  {author} {\bibinfo {author} {\bibfnamefont {J.~W.}\ \bibnamefont {Goodman}},\ }\href@noop {} {\emph {\bibinfo {title} {Statistical Optics}}}\ (\bibinfo  {publisher} {John Wiley \& Sons},\ \bibinfo {year} {2015})\BibitemShut {NoStop}%
\bibitem [{\citenamefont {Fyodorov}\ and\ \citenamefont {Safonova}(2023)}]{fyodorov2023intensity}%
  \BibitemOpen
  \bibfield  {author} {\bibinfo {author} {\bibfnamefont {Y.~V.}\ \bibnamefont {Fyodorov}}\ and\ \bibinfo {author} {\bibfnamefont {E.}~\bibnamefont {Safonova}},\ }\bibfield  {title} {\bibinfo {title} {Intensity statistics inside an open wave-chaotic cavity with broken time-reversal invariance},\ }\href@noop {} {\bibfield  {journal} {\bibinfo  {journal} {Phys. Rev. E}\ }\textbf {\bibinfo {volume} {108}},\ \bibinfo {pages} {044206} (\bibinfo {year} {2023})}\BibitemShut {NoStop}%
\bibitem [{\citenamefont {K{\"o}hnes}\ \emph {et~al.}(2026)\citenamefont {K{\"o}hnes}, \citenamefont {Che}, \citenamefont {Dietz},\ and\ \citenamefont {Guhr}}]{kohnes2026universality}%
  \BibitemOpen
  \bibfield  {author} {\bibinfo {author} {\bibfnamefont {S.}~\bibnamefont {K{\"o}hnes}}, \bibinfo {author} {\bibfnamefont {J.}~\bibnamefont {Che}}, \bibinfo {author} {\bibfnamefont {B.}~\bibnamefont {Dietz}},\ and\ \bibinfo {author} {\bibfnamefont {T.}~\bibnamefont {Guhr}},\ }\bibfield  {title} {\bibinfo {title} {Universality emerging in a universality: Derivation of the {Ericson} transition in stochastic quantum scattering and experimental validation},\ }\href@noop {} {\bibfield  {journal} {\bibinfo  {journal} {Phys. Rev. Lett.}\ }\textbf {\bibinfo {volume} {137}},\ \bibinfo {pages} {050403} (\bibinfo {year} {2026})}\BibitemShut {NoStop}%
\bibitem [{\citenamefont {Fyodorov}\ and\ \citenamefont {Savin}(2026)}]{fyodorov2026random}%
  \BibitemOpen
  \bibfield  {author} {\bibinfo {author} {\bibfnamefont {Y.~V.}\ \bibnamefont {Fyodorov}}\ and\ \bibinfo {author} {\bibfnamefont {D.~V.}\ \bibnamefont {Savin}},\ }\bibfield  {title} {\bibinfo {title} {Random matrix theory for chaotic wave scattering and transport},\ }\href@noop {} {\bibfield  {journal} {\bibinfo  {journal} {arXiv:2606.10957}\ } (\bibinfo {year} {2026})}\BibitemShut {NoStop}%
\bibitem [{\citenamefont {Yu}\ and\ \citenamefont {Zhang}(2002)}]{yu2002anti}%
  \BibitemOpen
  \bibfield  {author} {\bibinfo {author} {\bibfnamefont {Y.-K.}\ \bibnamefont {Yu}}\ and\ \bibinfo {author} {\bibfnamefont {Y.-C.}\ \bibnamefont {Zhang}},\ }\bibfield  {title} {\bibinfo {title} {On the anti-{W}ishart distribution},\ }\href@noop {} {\bibfield  {journal} {\bibinfo  {journal} {Physica A}\ }\textbf {\bibinfo {volume} {312}},\ \bibinfo {pages} {1} (\bibinfo {year} {2002})}\BibitemShut {NoStop}%
\bibitem [{\citenamefont {Janik}\ and\ \citenamefont {Nowak}(2003)}]{janik2003wishart}%
  \BibitemOpen
  \bibfield  {author} {\bibinfo {author} {\bibfnamefont {R.~A.}\ \bibnamefont {Janik}}\ and\ \bibinfo {author} {\bibfnamefont {M.~A.}\ \bibnamefont {Nowak}},\ }\bibfield  {title} {\bibinfo {title} {Wishart and {A}nti-{W}ishart random matrices},\ }\href@noop {} {\bibfield  {journal} {\bibinfo  {journal} {J. Phys. A: Math. Gen.}\ }\textbf {\bibinfo {volume} {36}},\ \bibinfo {pages} {3629} (\bibinfo {year} {2003})}\BibitemShut {NoStop}%
\bibitem [{\citenamefont {Vivo}\ \emph {et~al.}(2007)\citenamefont {Vivo}, \citenamefont {Majumdar},\ and\ \citenamefont {Bohigas}}]{vivo2007large}%
  \BibitemOpen
  \bibfield  {author} {\bibinfo {author} {\bibfnamefont {P.}~\bibnamefont {Vivo}}, \bibinfo {author} {\bibfnamefont {S.~N.}\ \bibnamefont {Majumdar}},\ and\ \bibinfo {author} {\bibfnamefont {O.}~\bibnamefont {Bohigas}},\ }\bibfield  {title} {\bibinfo {title} {Large deviations of the maximum eigenvalue in {W}ishart random matrices},\ }\href@noop {} {\bibfield  {journal} {\bibinfo  {journal} {J. Phys. A}\ }\textbf {\bibinfo {volume} {40}},\ \bibinfo {pages} {4317} (\bibinfo {year} {2007})}\BibitemShut {NoStop}%
\bibitem [{\citenamefont {Vivo}(2008)}]{vivo2008wishart}%
  \BibitemOpen
  \bibfield  {author} {\bibinfo {author} {\bibfnamefont {P.}~\bibnamefont {Vivo}},\ }\emph {\bibinfo {title} {From {W}ishart to {J}acobi ensembles: statistical properties and applications}},\ \href@noop {} {Ph.D. thesis},\ \bibinfo  {school} {Brunel University, School of Information Systems, Computing and Mathematics} (\bibinfo {year} {2008})\BibitemShut {NoStop}%
\bibitem [{\citenamefont {Majumdar}\ and\ \citenamefont {Schehr}(2024)}]{majumdar2024statistics}%
  \BibitemOpen
  \bibfield  {author} {\bibinfo {author} {\bibfnamefont {S.~N.}\ \bibnamefont {Majumdar}}\ and\ \bibinfo {author} {\bibfnamefont {G.}~\bibnamefont {Schehr}},\ }\href@noop {} {\emph {\bibinfo {title} {Statistics of Extremes and Records in Random Sequences}}}\ (\bibinfo  {publisher} {Oxford University Press},\ \bibinfo {year} {2024})\BibitemShut {NoStop}%
\bibitem [{\citenamefont {Dighe}\ \emph {et~al.}(2003)\citenamefont {Dighe}, \citenamefont {Mallik},\ and\ \citenamefont {Jamuar}}]{dighe2003analysis}%
  \BibitemOpen
  \bibfield  {author} {\bibinfo {author} {\bibfnamefont {P.~A.}\ \bibnamefont {Dighe}}, \bibinfo {author} {\bibfnamefont {R.~K.}\ \bibnamefont {Mallik}},\ and\ \bibinfo {author} {\bibfnamefont {S.~S.}\ \bibnamefont {Jamuar}},\ }\bibfield  {title} {\bibinfo {title} {Analysis of transmit-receive diversity in {Rayleigh} fading},\ }\href@noop {} {\bibfield  {journal} {\bibinfo  {journal} {IEEE Trans. Commun.}\ }\textbf {\bibinfo {volume} {51}},\ \bibinfo {pages} {694} (\bibinfo {year} {2003})}\BibitemShut {NoStop}%
\bibitem [{\citenamefont {Forrester}\ and\ \citenamefont {Kumar}(2019)}]{forrester2019recursion}%
  \BibitemOpen
  \bibfield  {author} {\bibinfo {author} {\bibfnamefont {P.~J.}\ \bibnamefont {Forrester}}\ and\ \bibinfo {author} {\bibfnamefont {S.}~\bibnamefont {Kumar}},\ }\bibfield  {title} {\bibinfo {title} {Recursion scheme for the largest $\beta$-{Wishart--Laguerre} eigenvalue and {Landauer} conductance in quantum transport},\ }\href@noop {} {\bibfield  {journal} {\bibinfo  {journal} {J. Phys. A: Math. Theor.}\ }\textbf {\bibinfo {volume} {52}},\ \bibinfo {pages} {42LT02} (\bibinfo {year} {2019})}\BibitemShut {NoStop}%
\bibitem [{\citenamefont {Tracy}\ and\ \citenamefont {Widom}(1994)}]{tracy1994level}%
  \BibitemOpen
  \bibfield  {author} {\bibinfo {author} {\bibfnamefont {C.~A.}\ \bibnamefont {Tracy}}\ and\ \bibinfo {author} {\bibfnamefont {H.}~\bibnamefont {Widom}},\ }\bibfield  {title} {\bibinfo {title} {Level-spacing distributions and the {Airy} kernel},\ }\href@noop {} {\bibfield  {journal} {\bibinfo  {journal} {Commun. Math. Phys.}\ }\textbf {\bibinfo {volume} {159}},\ \bibinfo {pages} {151} (\bibinfo {year} {1994})}\BibitemShut {NoStop}%
\bibitem [{\citenamefont {Johansson}(2000)}]{johansson2000shape}%
  \BibitemOpen
  \bibfield  {author} {\bibinfo {author} {\bibfnamefont {K.}~\bibnamefont {Johansson}},\ }\bibfield  {title} {\bibinfo {title} {Shape fluctuations and random matrices},\ }\href@noop {} {\bibfield  {journal} {\bibinfo  {journal} {Commun. Math. Phys.}\ }\textbf {\bibinfo {volume} {209}},\ \bibinfo {pages} {437} (\bibinfo {year} {2000})}\BibitemShut {NoStop}%
\bibitem [{\citenamefont {Johnstone}(2001)}]{johnstone2001distribution}%
  \BibitemOpen
  \bibfield  {author} {\bibinfo {author} {\bibfnamefont {I.~M.}\ \bibnamefont {Johnstone}},\ }\bibfield  {title} {\bibinfo {title} {On the distribution of the largest eigenvalue in principal components analysis},\ }\href@noop {} {\bibfield  {journal} {\bibinfo  {journal} {Ann. Stat.}\ }\textbf {\bibinfo {volume} {29}},\ \bibinfo {pages} {295} (\bibinfo {year} {2001})}\BibitemShut {NoStop}%
\bibitem [{\citenamefont {Majumdar}\ and\ \citenamefont {Schehr}(2014)}]{majumdar2014top}%
  \BibitemOpen
  \bibfield  {author} {\bibinfo {author} {\bibfnamefont {S.~N.}\ \bibnamefont {Majumdar}}\ and\ \bibinfo {author} {\bibfnamefont {G.}~\bibnamefont {Schehr}},\ }\bibfield  {title} {\bibinfo {title} {Top eigenvalue of a random matrix: large deviations and third order phase transition},\ }\href@noop {} {\bibfield  {journal} {\bibinfo  {journal} {J. Stat. Mech.: Theory Exp.}\ }\textbf {\bibinfo {volume} {2014}}\bibinfo  {number} { (1)},\ \bibinfo {pages} {P01012}}\BibitemShut {NoStop}%
\bibitem [{\citenamefont {Takeuchi}\ \emph {et~al.}(2011)\citenamefont {Takeuchi}, \citenamefont {Sano}, \citenamefont {Sasamoto},\ and\ \citenamefont {Spohn}}]{takeuchi2011growing}%
  \BibitemOpen
\bibfield  {number} {  }\bibfield  {author} {\bibinfo {author} {\bibfnamefont {K.~A.}\ \bibnamefont {Takeuchi}}, \bibinfo {author} {\bibfnamefont {M.}~\bibnamefont {Sano}}, \bibinfo {author} {\bibfnamefont {T.}~\bibnamefont {Sasamoto}},\ and\ \bibinfo {author} {\bibfnamefont {H.}~\bibnamefont {Spohn}},\ }\bibfield  {title} {\bibinfo {title} {Growing interfaces uncover universal fluctuations behind scale invariance},\ }\href@noop {} {\bibfield  {journal} {\bibinfo  {journal} {Sci. Rep.}\ }\textbf {\bibinfo {volume} {1}},\ \bibinfo {pages} {34} (\bibinfo {year} {2011})}\BibitemShut {NoStop}%
\bibitem [{\citenamefont {Fridman}\ \emph {et~al.}(2012)\citenamefont {Fridman}, \citenamefont {Pugatch}, \citenamefont {Nixon}, \citenamefont {Friesem},\ and\ \citenamefont {Davidson}}]{fridman2012measuring}%
  \BibitemOpen
  \bibfield  {author} {\bibinfo {author} {\bibfnamefont {M.}~\bibnamefont {Fridman}}, \bibinfo {author} {\bibfnamefont {R.}~\bibnamefont {Pugatch}}, \bibinfo {author} {\bibfnamefont {M.}~\bibnamefont {Nixon}}, \bibinfo {author} {\bibfnamefont {A.~A.}\ \bibnamefont {Friesem}},\ and\ \bibinfo {author} {\bibfnamefont {N.}~\bibnamefont {Davidson}},\ }\bibfield  {title} {\bibinfo {title} {Measuring maximal eigenvalue distribution of {Wishart} random matrices with coupled lasers},\ }\href@noop {} {\bibfield  {journal} {\bibinfo  {journal} {Phys. Rev. E}\ }\textbf {\bibinfo {volume} {85}},\ \bibinfo {pages} {020101} (\bibinfo {year} {2012})}\BibitemShut {NoStop}%
\bibitem [{\citenamefont {Makey}\ \emph {et~al.}(2020)\citenamefont {Makey}, \citenamefont {Galioglu}, \citenamefont {Ghaffari}, \citenamefont {Engin}, \citenamefont {Y{\i}ld{\i}r{\i}m}, \citenamefont {Yavuz}, \citenamefont {Bekta{\c{s}}}, \citenamefont {Nizam}, \citenamefont {Akbulut}, \citenamefont {{\c{S}}ahin} \emph {et~al.}}]{makey2020universality}%
  \BibitemOpen
  \bibfield  {author} {\bibinfo {author} {\bibfnamefont {G.}~\bibnamefont {Makey}}, \bibinfo {author} {\bibfnamefont {S.}~\bibnamefont {Galioglu}}, \bibinfo {author} {\bibfnamefont {R.}~\bibnamefont {Ghaffari}}, \bibinfo {author} {\bibfnamefont {E.~D.}\ \bibnamefont {Engin}}, \bibinfo {author} {\bibfnamefont {G.}~\bibnamefont {Y{\i}ld{\i}r{\i}m}}, \bibinfo {author} {\bibfnamefont {{\"O}.}~\bibnamefont {Yavuz}}, \bibinfo {author} {\bibfnamefont {O.}~\bibnamefont {Bekta{\c{s}}}}, \bibinfo {author} {\bibfnamefont {{\"U}.~S.}\ \bibnamefont {Nizam}}, \bibinfo {author} {\bibfnamefont {{\"O}.}~\bibnamefont {Akbulut}}, \bibinfo {author} {\bibfnamefont {{\"O}.}~\bibnamefont {{\c{S}}ahin}}, \emph {et~al.},\ }\bibfield  {title} {\bibinfo {title} {Universality of dissipative self-assembly from quantum dots to human cells},\ }\href@noop {} {\bibfield  {journal} {\bibinfo  {journal} {Nat. Phys.}\ }\textbf {\bibinfo {volume} {16}},\ \bibinfo {pages} {795} (\bibinfo {year} {2020})}\BibitemShut {NoStop}%
\bibitem [{\citenamefont {MacKinnon}(1985)}]{1985_MacKinnon}%
  \BibitemOpen
  \bibfield  {author} {\bibinfo {author} {\bibfnamefont {A.}~\bibnamefont {MacKinnon}},\ }\bibfield  {title} {\bibinfo {title} {The calculation of transport properties and density of states of disordered solids},\ }\href@noop {} {\bibfield  {journal} {\bibinfo  {journal} {Z. Phys. B}\ }\textbf {\bibinfo {volume} {59}},\ \bibinfo {pages} {385} (\bibinfo {year} {1985})}\BibitemShut {NoStop}%
\bibitem [{\citenamefont {Baranger}\ \emph {et~al.}(1991)\citenamefont {Baranger}, \citenamefont {DiVincenzo}, \citenamefont {Jalabert},\ and\ \citenamefont {Stone}}]{1991_Stone}%
  \BibitemOpen
  \bibfield  {author} {\bibinfo {author} {\bibfnamefont {H.~U.}\ \bibnamefont {Baranger}}, \bibinfo {author} {\bibfnamefont {D.~P.}\ \bibnamefont {DiVincenzo}}, \bibinfo {author} {\bibfnamefont {R.~A.}\ \bibnamefont {Jalabert}},\ and\ \bibinfo {author} {\bibfnamefont {A.~D.}\ \bibnamefont {Stone}},\ }\bibfield  {title} {\bibinfo {title} {Classical and quantum ballistic-transport anomalies in microjunctions},\ }\href@noop {} {\bibfield  {journal} {\bibinfo  {journal} {Phys. Rev. B}\ }\textbf {\bibinfo {volume} {44}},\ \bibinfo {pages} {637} (\bibinfo {year} {1991})}\BibitemShut {NoStop}%
\bibitem [{\citenamefont {Yu}\ \emph {et~al.}(2013)\citenamefont {Yu}, \citenamefont {Hillman}, \citenamefont {Choi}, \citenamefont {Lee}, \citenamefont {Feld}, \citenamefont {Dasari},\ and\ \citenamefont {Park}}]{2013_Park}%
  \BibitemOpen
  \bibfield  {author} {\bibinfo {author} {\bibfnamefont {H.}~\bibnamefont {Yu}}, \bibinfo {author} {\bibfnamefont {T.~R.}\ \bibnamefont {Hillman}}, \bibinfo {author} {\bibfnamefont {W.}~\bibnamefont {Choi}}, \bibinfo {author} {\bibfnamefont {J.~O.}\ \bibnamefont {Lee}}, \bibinfo {author} {\bibfnamefont {M.~S.}\ \bibnamefont {Feld}}, \bibinfo {author} {\bibfnamefont {R.~R.}\ \bibnamefont {Dasari}},\ and\ \bibinfo {author} {\bibfnamefont {Y.}~\bibnamefont {Park}},\ }\bibfield  {title} {\bibinfo {title} {Measuring large optical transmission matrices of disordered media},\ }\href@noop {} {\bibfield  {journal} {\bibinfo  {journal} {Phys. Rev. Lett.}\ }\textbf {\bibinfo {volume} {111}},\ \bibinfo {pages} {153902} (\bibinfo {year} {2013})}\BibitemShut {NoStop}%
\bibitem [{\citenamefont {Simon}\ and\ \citenamefont {Moustakas}(2004)}]{simon2004eigenvalue}%
  \BibitemOpen
  \bibfield  {author} {\bibinfo {author} {\bibfnamefont {S.~H.}\ \bibnamefont {Simon}}\ and\ \bibinfo {author} {\bibfnamefont {A.~L.}\ \bibnamefont {Moustakas}},\ }\bibfield  {title} {\bibinfo {title} {Eigenvalue density of correlated complex random {Wishart} matrices},\ }\href@noop {} {\bibfield  {journal} {\bibinfo  {journal} {Phys. Rev. E}\ }\textbf {\bibinfo {volume} {69}},\ \bibinfo {pages} {065101} (\bibinfo {year} {2004})}\BibitemShut {NoStop}%
\bibitem [{\citenamefont {Burda}\ \emph {et~al.}(2005)\citenamefont {Burda}, \citenamefont {Jurkiewicz},\ and\ \citenamefont {Wac{\l}aw}}]{burda2005spectral}%
  \BibitemOpen
  \bibfield  {author} {\bibinfo {author} {\bibfnamefont {Z.}~\bibnamefont {Burda}}, \bibinfo {author} {\bibfnamefont {J.}~\bibnamefont {Jurkiewicz}},\ and\ \bibinfo {author} {\bibfnamefont {B.}~\bibnamefont {Wac{\l}aw}},\ }\bibfield  {title} {\bibinfo {title} {Spectral moments of correlated {Wishart} matrices},\ }\href@noop {} {\bibfield  {journal} {\bibinfo  {journal} {Phys. Rev. E}\ }\textbf {\bibinfo {volume} {71}},\ \bibinfo {pages} {026111} (\bibinfo {year} {2005})}\BibitemShut {NoStop}%
\bibitem [{\citenamefont {Simon}\ \emph {et~al.}(2006)\citenamefont {Simon}, \citenamefont {Moustakas},\ and\ \citenamefont {Marinelli}}]{simon2006capacity}%
  \BibitemOpen
  \bibfield  {author} {\bibinfo {author} {\bibfnamefont {S.~H.}\ \bibnamefont {Simon}}, \bibinfo {author} {\bibfnamefont {A.~L.}\ \bibnamefont {Moustakas}},\ and\ \bibinfo {author} {\bibfnamefont {L.}~\bibnamefont {Marinelli}},\ }\bibfield  {title} {\bibinfo {title} {Capacity and character expansions: Moment-generating function and other exact results for {MIMO} correlated channels},\ }\href@noop {} {\bibfield  {journal} {\bibinfo  {journal} {IEEE Trans. Inf. Theory.}\ }\textbf {\bibinfo {volume} {52}},\ \bibinfo {pages} {5336} (\bibinfo {year} {2006})}\BibitemShut {NoStop}%
\bibitem [{\citenamefont {Zanella}\ \emph {et~al.}(2009)\citenamefont {Zanella}, \citenamefont {Chiani},\ and\ \citenamefont {Win}}]{zanella2009marginal}%
  \BibitemOpen
  \bibfield  {author} {\bibinfo {author} {\bibfnamefont {A.}~\bibnamefont {Zanella}}, \bibinfo {author} {\bibfnamefont {M.}~\bibnamefont {Chiani}},\ and\ \bibinfo {author} {\bibfnamefont {M.~Z.}\ \bibnamefont {Win}},\ }\bibfield  {title} {\bibinfo {title} {On the marginal distribution of the eigenvalues of {Wishart} matrices},\ }\href@noop {} {\bibfield  {journal} {\bibinfo  {journal} {IEEE Trans. Commun.}\ }\textbf {\bibinfo {volume} {57}},\ \bibinfo {pages} {1050} (\bibinfo {year} {2009})}\BibitemShut {NoStop}%
\bibitem [{\citenamefont {Gaspard}\ and\ \citenamefont {Goetschy}(2025{\natexlab{a}})}]{gaspard2025radiant}%
  \BibitemOpen
  \bibfield  {author} {\bibinfo {author} {\bibfnamefont {D.}~\bibnamefont {Gaspard}}\ and\ \bibinfo {author} {\bibfnamefont {A.}~\bibnamefont {Goetschy}},\ }\bibfield  {title} {\bibinfo {title} {Radiant field theory: A transport approach to shaped wave transmission through disordered media},\ }\href@noop {} {\bibfield  {journal} {\bibinfo  {journal} {Phys. Rev. Lett.}\ }\textbf {\bibinfo {volume} {135}},\ \bibinfo {pages} {033804} (\bibinfo {year} {2025}{\natexlab{a}})}\BibitemShut {NoStop}%
\bibitem [{\citenamefont {Gaspard}\ and\ \citenamefont {Goetschy}(2025{\natexlab{b}})}]{gaspard2025transmission}%
  \BibitemOpen
  \bibfield  {author} {\bibinfo {author} {\bibfnamefont {D.}~\bibnamefont {Gaspard}}\ and\ \bibinfo {author} {\bibfnamefont {A.}~\bibnamefont {Goetschy}},\ }\bibfield  {title} {\bibinfo {title} {Transmission eigenvalue distribution in disordered media from radiant field theory},\ }\href@noop {} {\bibfield  {journal} {\bibinfo  {journal} {Phys. Rev. Research}\ }\textbf {\bibinfo {volume} {7}},\ \bibinfo {pages} {033071} (\bibinfo {year} {2025}{\natexlab{b}})}\BibitemShut {NoStop}%
\end{thebibliography}%

\end{document}